\documentclass[preprint,preprintnumbers,amsmath,amssymb,floatfix,nofootinbib,superscriptaddress]{revtex4}
\usepackage[dvips]{graphicx}
\usepackage{url}
\usepackage{subfigure}
\usepackage{color}

\newcommand{\as}{\alpha_s}

\def\be{\begin{equation}}
\def\ee{\end{equation}}
\newcommand{\ba}{\begin{eqnarray}}
\newcommand{\ea}{\end{eqnarray}}

\def\qb{\bar q}
\def\Qb{\bar Q}
\def\tb{\bar t}
\def\bb{\bar b}
\def\gam{\gamma}
\def\pb{\bar p}
\def\oo{\mathcal{O}}

\def\as{\alpha_s}
\def\nn{\nonumber}
\def\lo{\mathrm{LO}}
\def\nlo{\mathrm{NLO}}

\def\nlf{n_{\mathrm{lf}}}

\def\ln{\mathrm{ln}}
\def\q5{5\mathrm{q}}
\def\gh5{5\mathrm{gh}}
\def\spa#1.#2{\left\langle#1#2\right\rangle}
\def\spb#1.#2{\left[#1#2\right]}

\def\muR{\mu_R}
\def\muF{\mu_F}

\def\spa#1.#2{\langle#1#2\rangle}
\def\spb#1.#2{[#1#2]}
\def\spab#1.#2.#3{\langle#1|#2|#3]}
\def\spba#1.#2.#3{[#1|#2|#3\rangle}
\def\spaa#1.#2.#3{\langle#1|#2|#3\rangle}
\def\spbb#1.#2.#3{[#1|#2|#3]}
\def\spaxa#1.#2.#3.#4{\langle#1|#2|#3|#4\rangle}
\def\spbxb#1.#2.#3.#4{[#1|#2|#3|#4]}

\begin{document}          
%
%
\title{Hard-photon production with $b$ jets at hadron colliders}
\author{H.~B.~Hartanto}
\email{hartanto@physik.rwth-aachen.de}
\affiliation{Insitut f\"{u}r Theoretische Teilchenphysik und Kosmologie,
RWTH Aachen University, D-52056 Aachen, Germany}
\affiliation{Physics Department, Florida State University,
Tallahassee, FL 32306-4350, USA}
\author{L.~Reina}
\email{reina@hep.fsu.edu}
\affiliation{Physics Department, Florida State University,
Tallahassee, FL 32306-4350, USA}

\date{\today}

\begin{abstract}
  We present total and differential cross sections for the production
  of a hard photon with up to two $b$ jets at both the Tevatron with
  center-of-mass energy 1.96~TeV and the Large Hadron Collider with
  center-of-mass energy 8~TeV, including Next-to-Leading Order (NLO)
  QCD corrections and full $b$-quark mass effects. We study the
  theoretical uncertainty due to the residual renormalization- and
  factorization-scale dependence and explain its origin on the basis
  of the different subprocesses contributing to the NLO cross section.
  We specifically address the case of the production of a hard photon
  with at least one $b$ jet and compare the NLO QCD predictions in
  both the Four- and Five-Flavor-Number Schemes to the experimental
  measurements obtained by CDF and D0.
\end{abstract}
%
\maketitle

\section{Introduction}
\label{sec:intro}

The associated production of a hard photon with a heavy-quark pair
($Q\bar{Q}\gamma$ for $Q=t,b$) plays a very important role in the
physics of both the Tevatron and the Large Hadron Collider (LHC) 
to the extent it provides direct information on the third-generation
quark electromagnetic couplings and the bottom-quark parton density.

A measurement of $t\bar{t}\gamma$ production can provide a direct test
of the $t\bar{t}\gamma$ coupling provided suitable selection cuts can
isolate the emission of a hard photon from the produced top-quark
pair~\cite{Baur:2004uw,Baur:2006ck}.  Next-to-Leading Order (NLO) QCD
corrections to this process have been calculated for on-shell top
quarks~\cite{Duan:2009zza} and for off-shell top
quarks~\cite{Melnikov:2011ta}, where NLO QCD corrections for the decay
process have also been taken into account using a narrow-width
approximation, and the problem of distinguishing hard-photon emission
from production and decay has been addressed.

On the other hand, the associated production of a photon with a
$b\bar{b}$ pair is a crucial component of the theoretical prediction
for direct-photon production with $b$ jets which, once compared with
experiments, will provide a direct access to the bottom-quark parton
density in nucleons and help understanding the nature of the $b$-quark
parton distribution function (PDF).  Several subtle issues enter the
comparison of theoretical predictions with existing and future
experimental measurements and progress is still needed to be able to
directly constrain the $b$-quark parton distribution function and
investigate the presence of an intrinsic $b$-quark density in nucleons
as opposed to a purely perturbative $b$-quark density obtained from the
evolution of the gluon parton density. Constraining and understanding
the $b$-quark parton density will play a very important role in
improving the accuracy with which other crucial processes like the
associated production with weak gauge bosons ($W/Z+b$), a background
to Higgs production, and the associated production with a
scalar/pseudoscalar ($H/A+b$), a clear signal of new physics, can be
predicted.

\begin{figure}[t!]
\begin{center}
\includegraphics[width=8cm]{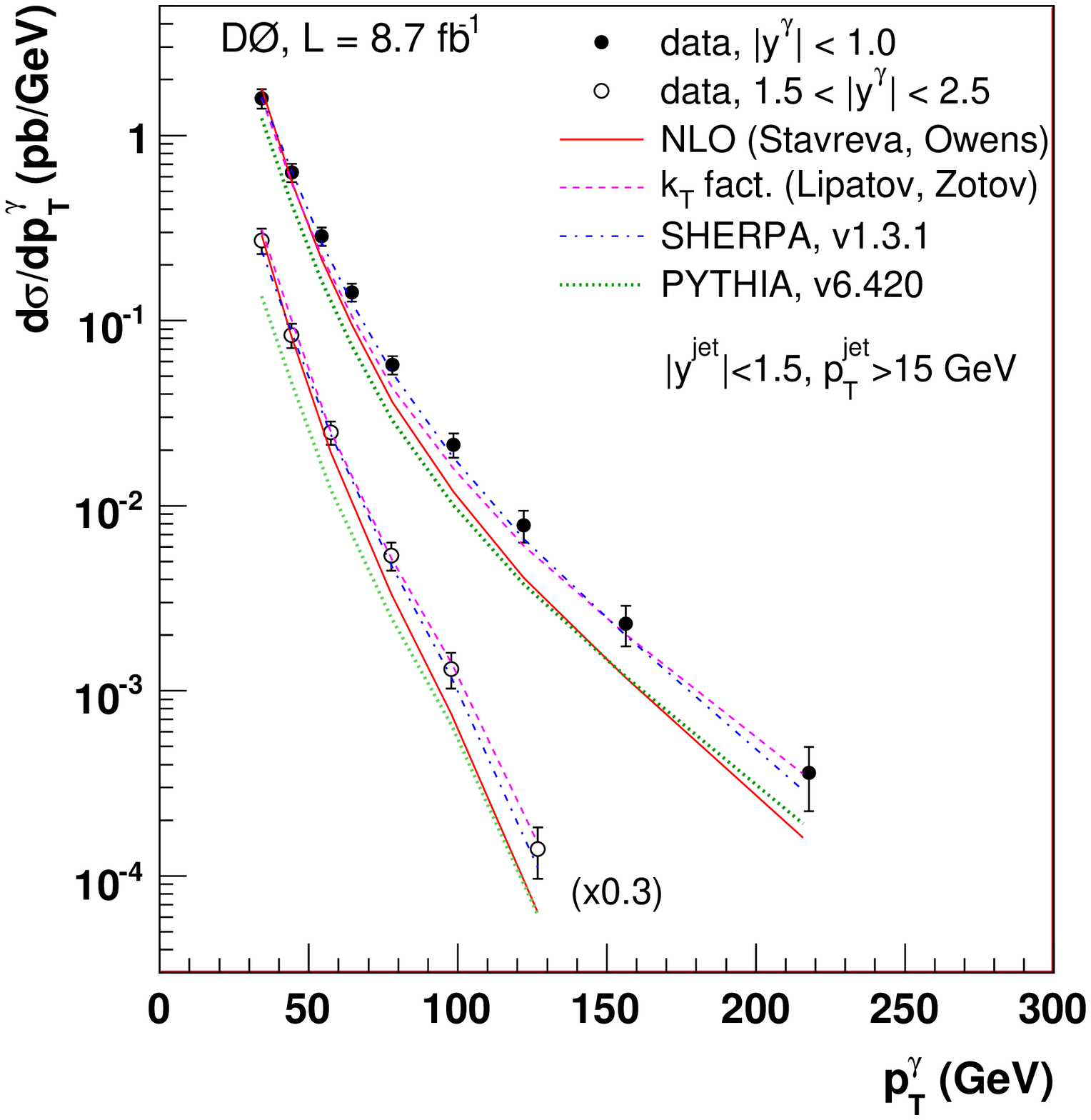}
\includegraphics[width=7cm]{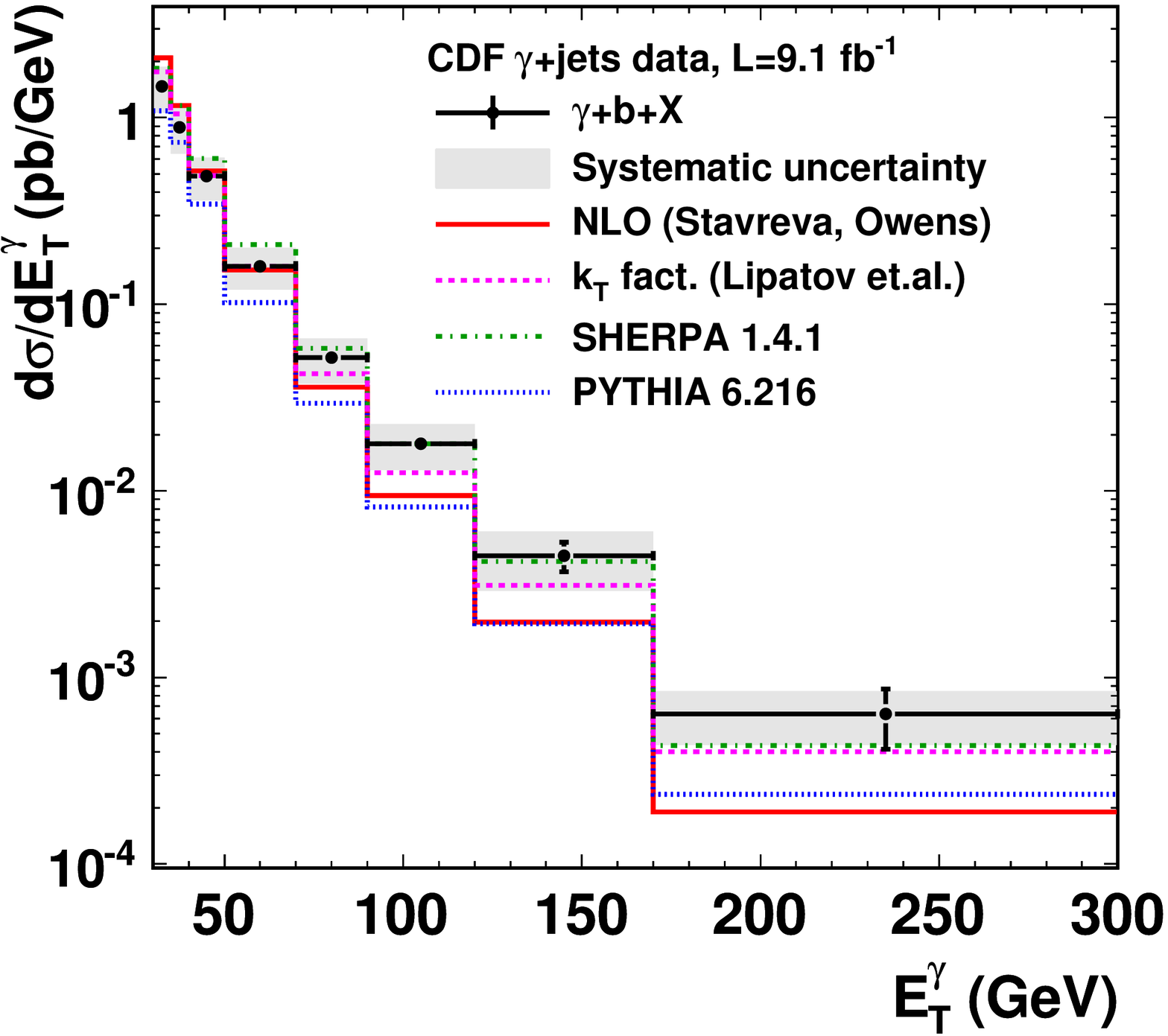}
\caption{The photon transverse-momentum distribution measured by the
  D0 \cite{Abazov:2012ea} (left) and CDF \cite{Aaltonen:2013coa}
  (right) collaborations for $p\pb \rightarrow \gam+b+X$ process at
  the Tevatron in comparison with theoretical predictions.}
\label{fig:bbgamma_exp}
\end{center}
\end{figure}

The NLO prediction for direct photon production in association with
one $b$ jet has been calculated in the so-called
\textit{variable-flavor scheme} (VFS) or \textit{five-flavor-number
scheme} (5FNS)~\cite{Stavreva:2009vi}, where a $b$-quark parton
density is assumed in the initial state and the $b$ quark is treated
as massless.  On the experimental side, the $\gam+b+X$ process has
been measured at the Tevatron by the D0 collaboration with 1
fb$^{-1}$~\cite{Abazov:2009de} and 8.7 fb$^{-1}$~\cite{Abazov:2012ea}
data sets as well as by the CDF collaboration with 86
pb$^{-1}$~\cite{Aaltonen:2009wc} and more recently 9.1
fb$^{-1}$~\cite{Aaltonen:2013coa} data sets.  The $p_T(\gam)$
distributions for the $p\pb \rightarrow \gam+b+X$ process at the
Tevatron from the most recent D0 \cite{Abazov:2012ea} and 
CDF~\cite{Aaltonen:2013coa} results are shown in
Fig.~\ref{fig:bbgamma_exp}. The experimental data are compared with
the predictions from the VFS/5FNS NLO calculation in
\cite{Stavreva:2009vi}, as well as other predictions from
Pythia~\cite{Sjostrand:2006za,Sjostrand:2007gs},
Sherpa~\cite{Gleisberg:2003xi,Gleisberg:2008ta}, and a
calculation which uses the $k_T$-factorization approach
\cite{Lipatov:2012rg} that contains only partial NLO corrections but
selected higher-order effects.  In the intermediate to high photon
transverse-momentum region, one notices some discrepancies between the
data and the VFS/5FNS NLO calculation.

In this paper we present the NLO QCD results for hard-photon
production with either one or two $b$ jets, using a
\textit{fixed-flavor scheme} (FFS) or \textit{four-flavor-number
scheme} (4FNS).  The calculation consists of the NLO QCD corrections
to $pp(p\pb) \rightarrow b\bb\gam$, where the $b$ quark is treated as
massive and no $b$-quark parton density is assumed in the initial
state\footnote{We notice that the production of an off-shell photon in
association with a $b\bar{b}$ pair at NLO in QCD has been presented in
Ref.~\cite{Hirschi:2011pa}, but cannot be used for on-shell direct
photon production}. Details of the FFS/4FNS calculation are presented
in Section~\ref{sec:nlo_2b_1b} , while various kinematic distributions
for both the $\gamma+b$ and $\gamma+2b$ jets are presented in
Section~\ref{sec:results}, including their theoretical systematic
uncertainty.  The comparison with CDF and D0
data~\cite{Abazov:2012ea,Aaltonen:2013coa}, as well as other existing
theoretical results~\cite{Stavreva:2009vi}, is also discussed in
Section~\ref{sec:results}.  Finally, Section~\ref{sec:conclusions}
summarizes our conclusions and suggests possible future developments.

\section{NLO QCD corrections to $\gamma+2b$ and $\gamma+1b$ jets}
\label{sec:nlo_2b_1b}

In this section we present the most relevant aspects of the
calculation of the $\oo(\alpha_s)$ corrections to $pp(p\pb)\rightarrow
Q\bar{Q}\gamma$ for $Q=t,b$. The structure of the calculation and the
techniques used in its realization are summarized in
Sec.~\ref{subsec:nlo_virt_real}. A complete discussion of the details
can be found in \cite{bayu_thesis}.  As briefly mentioned in
Sec.~\ref{sec:intro}, the case of hard-photon production with one $b$
jet can be addressed using both a 5FNS and a 4FNS approach. We
will discuss the comparison between the two approaches in more detail
in Sec.~\ref{subsec:1b_4v5_theory}.  Finally, we will introduce and
discuss the impact of different choices of photon-isolation
prescription in Sec.~\ref{subsec:photon_isolation}.

\subsection{Structure of $\oo(\alpha_s)$ corrections}
\label{subsec:nlo_virt_real}

\begin{figure}
\begin{center}
\includegraphics[width=15cm]{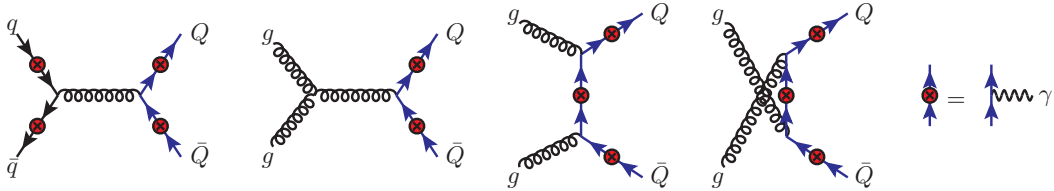}
\vspace{-0.5cm}
\caption{Tree-level Feynman diagrams for $pp(p\pb) \rightarrow
Q\Qb\gam$ production, corresponding to the $q\qb\rightarrow Q\Qb\gam$
and $gg\rightarrow Q\Qb\gam$ subprocesses. For each diagram, the red
circled crosses correspond to all possible photon insertions.}
\label{fig:4fns_tree_level}
\end{center}
\end{figure}

At tree level the $pp(p\pb)\rightarrow Q\bar{Q}\gamma$ ($Q=t,b$)
process consists of two partonic subprocesses, namely
$q\bar{q}\rightarrow Q\bar{Q}\gamma$ and $gg\rightarrow
Q\bar{Q}\gamma$, as illustrated in Fig.~\ref{fig:4fns_tree_level}.
The first order of QCD corrections consist of $\oo(\alpha_s)$ virtual
one-loop corrections to the tree-level subprocesses as well as
$\oo(\alpha_s)$ real corrections in the form of three subprocesses
with an additional radiated parton, namely $q\bar{q}\rightarrow
Q\bar{Q}\gamma+g$, $gg\rightarrow Q\bar{Q}\gamma+g$, and
$gq(g\bar{q})\rightarrow Q\bar{Q}\gamma+q(\bar{q})$.

We have calculated the $\oo(\alpha_s)$ virtual one-loop corrections
using a Feynman-diagram approach and two independent calculations,
based on an in-house code and on the NLOX package~\cite{Reina:2011mb}
respectively.  Both codes relies on the FORM symbolic
manipulation program \cite{Vermaseren:2000nd} 
to decompose tensor integrals in terms of tensor-integral coefficients
and spinor structures, as well as to interfere the one-loop and
tree-level amplitudes. 
Tensor one-loop integrals are reduced to a linear combination of
scalar integrals using various techniques such as the
Passarino-Veltman (PV)~\cite{Passarino:1978jh},
Denner-Dittmaier~\cite{Denner:2005nn}, and Diakonidis et
al.~\cite{Diakonidis:2008ij} methods.  One-loop scalar integrals are
evaluated using the QCDLoop package \cite{Ellis:2007qk}.

UV and IR divergences have been extracted using dimensional
regularization. UV divergences arising from self-energy and vertex
diagrams are cancelled by introducing counterterms for the external
fields, the strong coupling, and the heavy-quark mass. The QED
coupling does not renormalize at the first order in $\alpha_s$.  IR
divergences arising in vertex, box, and pentagon diagrams are
cancelled by analogous IR divergences in the real-emission part of the
NLO cross section.  A detailed discussion of the UV and IR divergences
of the $(q\qb,gg) \rightarrow Q\Qb\gam$ virtual amplitudes is given in
Ref.~\cite{bayu_thesis}. 

Real-emission corrections have been computed using a
phase-space-slicing method with both a soft ($\delta_s$) and a
collinear ($\delta_c$) cutoffs to isolate and compute the IR singular
terms of the cross section and the corresponding finite contributions.
The cancellation of IR singularities between virtual and real
corrections has been verified and the independence of the physical
cross section of the choice of $\delta_s$ and $\delta_c$ has been
thoroughly proved~\cite{bayu_thesis}.

On top of internal independent cross checks, we have also interfaced
our routines for the one-loop virtual corrections with the
Sherpa~\cite{Gleisberg:2003xi,Gleisberg:2008ta} Monte Carlo
event generator, which implements the Catani-Seymour 
dipole-subtraction formalism~\cite{Catani:1996vz,Catani:2002hc}, and
found agreement at the level of the partonic NLO cross sections.

\subsection{4FNS vs 5FNS}
\label{subsec:1b_4v5_theory}

The NLO QCD calculation of $Q\Qb\gam$ hadronic production allows us to
study the phenomenology of both $t\tb\gam$ and $b\bb\gam$ production
at the Tevatron and the LHC.  Practically, one can simply specify
the mass ($m_Q$ = $m_t$ or $m_b$) as well as the charge of the heavy
quark ($Q_Q$ = $Q_t$ or $Q_b$) to switch from one to the other.  The
case of a final-state bottom-quark pair, however, requires some extra
care due to both theoretical and experimental issues. 

$t\bar{t}\gam$ is calculated assuming five massless quark flavors,
i.e.  in a 5FNS. The short lifetime of the top quark allows it to
decay (dominantly via $t \rightarrow b W$) before it hadronizes. In
calculating inclusive observables in $t\tb\gam$ production, the top
quark can be considered as a stable final state, as done in
\cite{Duan:2009zza} and \cite{Melnikov:2011ta}. Alternatively, one can
consider more exclusive modes, where the decay of the top-quark pair
is also explicitly accounted for.  The study of $t\tb\gam$ production
including NLO QCD corrections both in the production and decay stages
is done in \cite{Melnikov:2011ta}. We have reproduced results for a
stable $t\bar{t}$ pair and found full agreement with
Ref.~\cite{Melnikov:2011ta}, for the same setup of external
parameters\footnote{Full details can be found in
\cite{bayu_thesis}}. Since a comparison with Ref.~\cite{Duan:2009zza}
had already been presented in Ref.~\cite{Melnikov:2011ta}, we have not
investigated it any further in this context.

\begin{figure}
\begin{center}
\includegraphics[width=13cm]{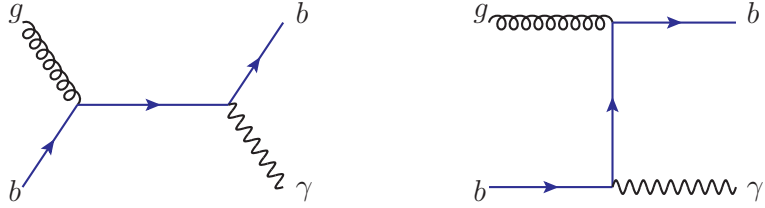}
\caption{Tree-level Feynman diagrams for   
$pp(p\pb) \rightarrow b(\bb)\gam+X$ production in the 5FNS,
  corresponding to the $gb\rightarrow b\gam$ subprocess.}
\label{fig:5fns_tree_level}
\end{center}
\end{figure}

On the other hand, a bottom quark in the final state will form a jet
that can be detected experimentally via $b$-tagging, i.e. imposing
specific cuts on some $b$-jet kinematic variables, typically its
transverse momentum and pseudorapidity. In this paper we consider the
following cases:
\begin{itemize}
\item  at least two $b$ jets observed in the final state ($pp(p\pb)
  \rightarrow b\bb\gam + X$, ``$2b$-tag"), 
\item at least one $b$ jet observed in the final state ($pp(p\pb)
  \rightarrow b(\bb)\gam + X$, ``$1b$-tag"),
\end{itemize}
The NLO QCD calculation of the $2b$-tag case can only proceed in the
4FNS, where one assumes only four massless quark flavors, the $b$
quark is treated as massive, and does not appear in the initial state.
This is exactly the $Q\bar{Q}\gam$ calculation presented in this paper
with $Q=b$ ($m_Q=m_b$, $Q_Q=Q_b$). On the other hand, the NLO QCD
calculation of the $1b$-tag case can be done using both the 4FNS and
the 5FNS, where instead the $b$ flavor is treated as massless and an
initial state $b$-quark density is introduced. The tree level
processes are different and are given in
Fig.~\ref{fig:4fns_tree_level} for the 4FNS and in
Fig.~\ref{fig:5fns_tree_level} for the 5FNS.
The $\oo(\alpha_s)$ corrections are then calculated as one-loop
virtual and real-emission corrections to the corresponding tree level
processes. The 4FNS case is already discussed in
Sec.~\ref{subsec:nlo_virt_real}. For the 5FNS case we list the
processes entering the NLO QCD calculation and their role in
Table~\ref{table:5FNSsubproc}.
\begin{table}
\begin{center}
\caption{List of subprocesses that contribute to the LO and NLO VFS/5FNS
  calculation of $pp(p\pb) \rightarrow \gam b + X$ process, with
  $i=q,\qb$ and $Q=b,\bb$.}  \renewcommand{\arraystretch}{1.5}
\begin{tabular}{ | c | c | }
  \hline
  \hline
  Part & Subprocess \\
  \hline
  LO and NLO virtual & $ Qg \rightarrow \gam Q $ \\
  \hline
  NLO real &  $ Qg \rightarrow \gam Q g $ \\
  NLO real &  $ iQ \rightarrow \gam Q i $ \\
  NLO real &  $ QQ \rightarrow \gam Q Q $ \\
  NLO real &  $ q\qb \rightarrow \gam b \bb $ \\
  NLO real &  $ gg \rightarrow \gam b \bb $ \\
  \hline
  \hline
\end{tabular}
\end{center}
\label{table:5FNSsubproc}
\end{table}
In both cases one requires that at least one $b$ jet is
reconstructed and tagged in the final state. In the 4FNS, the selected
sample will of course also include events with two $b$ jets, as well
as events with one or two $b$ jets and a light jet.

As it is well known, the 5FNS approach naturally arise from the 4FNS
calculation when one considers that the integration over the phase
space of the final-state untagged $b$ quark generates logarithms of
the form $\ln(Q/m_b)$, where technically $Q$ is the upper bound on the
$p_T$ of the unobserved $b$-quark. For large $Q$ ($Q\gg m_b$) these
logarithms can become large and spoil the convergence of the
perturbative expansion of the cross section.  These logarithms however
can be factored out and resummed using renormalization-group arguments
in the form of DGLAP equations by introducing a bottom-quark PDF,
\begin{equation}
f_b^{p/\pb}(x,\mu) = \frac{\as(\mu)}{\pi} \ln\bigg(\frac{Q}{m_b}\bigg) \int_x^1 \frac{dy}{y}
P_{gq}\bigg(\frac{x}{y}\bigg) f_g^{p/\pb}(x,\mu),
\label{eq:bpdf}
\end{equation}
where $f_g^{p/\pb}(x,\mu)$ is the gluon PDF and $P_{gq}$ is the
Altarelli-Parisi splitting function for $g \rightarrow q\qb$.  By
defining the $b$-quark PDF, the 5FNS approach restructures the
calculation as an expansion in terms of $\as$ and the potentially
large $\ln(Q/m_b)$ logarithms.  As a result of the 5FNS approach, the
process where at least one $b$ jet is identified in the final state
now starts at LO with the $gb \rightarrow b\gam$ subprocesses shown in
Fig.~\ref{fig:5fns_tree_level}, with $m_b = 0$. If the 5FNS approach
emphasizes the role of these \textit{initial-state} logarithms, and
add stability to the theoretical results by resumming leading and
subleading families of such logarithms, it nevertheless neglects other
contributions that do not appear in the 5FNS calculation (as one can
notice by comparing the list of subprocesses and their $\oo(\alpha_s)$
corrections) because they are not directly affected by the resummation
logarithms. At the same time, different subprocesses enter at
different perturbative orders in the 4FNS or 5FNS calculation of the
same hadronic process. For instance, in the case of
$pp(p\pb)\rightarrow \gam+b+X$, the $q\bar{q}\rightarrow b\bar{b}\gam$
subprocess enters at lowest order in the 4FNS calculation, while it
counts as an $\oo(\alpha_s)$ correction in the 5FNS calculation. As
such, it enters the 5FNS calculation as a tree-level process, while it
is included in the 4FNS calculation together with its own
$\oo(\alpha_s)$ corrections. If a particular subprocess, as it is the
case for $q\bar{q}\rightarrow b\bar{b}\gam$, dominates in a given
kinematic regime or at a given center-of-mass energy, including it at
tree-level or at the one-loop level can make a drastic difference both
quantitatively (one-loop corrections may be large) and qualitatively
(NLO corrections in general reduce the theoretical systematic
uncertainty from renormalization- and factorization-scale
dependence). Therefore, even if 4FNS and 5FNS approaches are just two
different ways of reorganizing the QCD perturbative expansion of a
given physical observable, they may show significantly different
behaviors within the first few orders of the perturbative
expansion. Moreover, being the logarithms resummed in the 5FNS
approach of fundamental kinematic nature, the interpretation of the
comparison between the 4FNS and 5FNS approaches may require to look at
both total and differential cross sections.  In
Sec.~\ref{subsec:bbgam_4v5} we will quantitatively illustrate the
comparison between the two different approaches and in
Sec.~\ref{subsec:1bgam_vs_tevdata} we will discuss the comparison of
both predictions with both CDF and D0 data.

\subsection{Photon isolation}
\label{subsec:photon_isolation}

Photons in a hadronic environment are usually distinguished into
\textit{prompt photons}, when they are directly produced in the hard
interaction, and \textit{secondary photons}, when they originate from
the hadronization phase of a hadronic jet or the decay of unstable
hadrons (e.g. $\pi^0\rightarrow\gam\gam$). While the production of
prompt photons can be described in perturbation theory, the production
of secondary photons can only be modeled and can therefore introduce a
large parametric uncertainty in any given calculation.  Since
secondary photons tend to preeminently occur in regions of the
detector with abundant hadronic activity, in particular within or
close to jets, their effect can be eliminated by imposing so-called
\textit{isolation cuts} which specifically limit the hadronic activity
around a given photon. Prompt photons become then \textit{isolated
  photons} and can be easily disentangled. 

The main theoretical caveat in implementing a given prescription to
\textit{isolate} prompt photons from the hard interaction is that such
procedure can veto regions of phase space responsible for soft QCD
radiation and could therefore spoil the cancellation of infrared
divergences between virtual and real corrections in a perturbative
QCD calculation. As soon as some residual hadronic activity is
admitted in the region around the photon, very energetic collinear
final-state partons can produce a small parton-photon invariant mass
and the corresponding collinear divergences need therefore to be
reabsorbed into suitable fragmentation functions\footnote{In our
  calculation this happens in the $qg\rightarrow Q\bar{Q}\gam+q$
  channel when the photon becomes collinear to the massless final
  state quark. Notice that the hard-photon cut that is imposed on the
  transverse momentum of the photon (see
  Sec.~\ref{subsec:results_setup}) eliminates initial-state
  parton-photon singularities and all soft-photon singularities.}. 
To extract the quark-photon final state collinear singularity
encountered in our calculation, we have also used the phase-space
slicing method. The cross section for prompt-photon production is then given by,
\begin{equation}
\sigma^\gam (\muR,\muF,M_F) = \sigma_{\mathrm{direct}}^\gam(\muR,\muF) + \int_0^1 dz \sum_i
\sigma_i(\muR,\muF,M_F) D_{i \rightarrow \gam} (z,M_F),
\label{eq:photonxsec}
\end{equation}
where $\sigma_{\mathrm{direct}}^\gam$ represents the cross section for
the direct component while $\sigma_i$ denotes the cross section for
the production of a parton $i$ that further fragments into a photon.
The probability for a parton $i$ to fragment into a photon is
represented by the corresponding photon fragmentation functions (FFs),
$D_{i\rightarrow \gam}(z,M_F)$, where $z$ is the fraction of the
parton momentum that is carried by the photon, and $M_F$ is the
fragmentation scale. Examples of available FFs in the literature are
by Bourhis, Fontannaz and Guillet (set I and II) \cite{Bourhis:1997yu}
and by Gehrmann-de Ridder and Glover \cite{GehrmannDeRidder:1998ba}.
Fragmentation functions for final-state partons, like parton
distribution functions for initial-state partons, are intrinsically
non perturbative and introduce into the calculation the same kind of
uncertainty in the modeling of secondary photons that one originally
wanted to eliminate. How relevant the contribution of fragmentation
functions is depends on the chosen isolation prescription.

In our calculation we used two main prescriptions that we denote as
\textit{fixed-cone} and \textit{smooth-cone} prescriptions. The
\textit{fixed-cone} prescription is commonly used in experiments and
limits the hadronic activity inside a cone of
radius $R_0$ around the photon by imposing that the hadronic
transverse energy inside the cone does not exceed a maximum value,
 $E_T^{\mathrm{max}}$, set by the experiment, i.e.
\begin{equation}
\sum_{\in R_0} E_T(\mathrm{had}) < E_T^{\mathrm{max}}\,\,\,,
\label{eq:photoiso}
\end{equation}
where $R_0 = \sqrt{\Delta \eta^2 + \Delta \phi^2}$, and $\Delta \eta$ and
$\Delta\phi$ are the pseudorapidity and azimuthal angle differences
between the photon and a jet.  After the isolation cut, the value of $z$ is
typically large, and since the FFs are dominant in the low $z$ region,
the isolation procedure suppresses the fragmentation contribution
substantially.

Alternatively, the \textit{smooth-cone} isolation prescription
introduced in Ref.~\cite{Frixione:1998jh} limits the hadronic activity
around a photon by imposing a threshold on the transverse
hadronic energy within a cone about the photon that varies with the
radial distance from the photon, i.e
\begin{equation} 
\sum_{i}  E_T^{i} \, \theta(R-R_{{i},\gamma})
 < \epsilon E_T^{\gamma}\bigg(\frac{1-\cos{R}}{1-\cos{R_0}}\bigg)
 \qquad \mbox{for all} \quad R \le R_0,
\label{eq:frixiso}
\end{equation}
where the $i$ summation runs over all final-state partons in the
process and $E_T^{i(\gam)}$ is the transverse energy of the parton
(photon). $R_0$ is the size of the isolation cone, $\epsilon$ is an
isolation parameter of $O(1)$, and
\begin{equation}
R_{i,\gam}=\sqrt{(\Delta \eta_{i,\gam})^2 + (\Delta \phi_{i,\gam})^2}. \nn
\end{equation}
The $\theta$-function ensures that the $i$ summation only receives
contributions from partons that lie inside the isolation cone. $R =
R_{i,\gam}$ if there is only one parton inside the isolation cone,
while for the case where more than one parton is present inside the
cone, $R$ is the largest $R_{i,\gam}$ inside the cone.  The r.h.s of
Eq.~\ref{eq:frixiso} vanishes as $R\rightarrow 0$, thus the collinear
configurations are suppressed while soft radiation is allowed to
be present arbitrarily close to the photon.  Since the collinear
configurations are completely removed, there is no fragmentation
component in Eq.~\ref{eq:photonxsec}.

The fragmentation contribution in the $pp(p\pb) \rightarrow b\bb\gam$
calculation is included at $\oo(\alpha\as^2)$. Due to the photon
isolation requirement, a photon cannot fragment from the tagged
$b/\bb$ quark.  In the $2b$-tag case (as well as in $t\tb\gam$
production), the photon can only fragment off a light parton $j$,
i.e. $\sigma_i$ in Eq.~\ref{eq:photonxsec} is the cross section for
the $pp(p\pb) \rightarrow b\bb j$ process calculated at LO, ($\sigma_i
= \sigma_{LO}(pp(p\pb) \rightarrow b\bb j)$).  We notice that
$\sigma_{LO}(pp(p\pb) \rightarrow b\bb j)$ is finite since we impose a
cut on the photon transverse momentum. For the $1b$-tag case, in
addition to the same contribution present in the $2b$-tag case, the
photon can also fragment off an unidentified $b/\bb$ quark.  The LO
$pp(p\pb) \rightarrow b\bb j$ cross section is divergent in this case
since the light parton in the final state can be soft and/or
collinear. To overcome this problem, we should start from the
$pp(p\pb) \rightarrow b\bb$ cross section at NLO in QCD.  We have
implemented the $\oo(\as)$ real corrections to $pp(p\pb) \rightarrow
b\bb$ using a phase-space slicing method with two cutoffs, while we
have taken the $\oo(\as)$ virtual corrections from the MCFM package
\cite{Campbell:2010ff}.  We also notice that when the photon is
fragmented off of a $b/\bb$ quark, terms proportional to
$\ln(M_F^2/m_b^2)$, arising from the collinear configuration of the
$b\rightarrow b\gam$ splitting in the (direct) $pp(p\pb) \rightarrow
b\bb\gam$ process, have to be subtracted to avoid double counting
since those terms have been included and resummed in the
$b$ quark-to-photon fragmentation function via DGLAP evolution equations.

\section{Results}
\label{sec:results}

In this Section we present numerical results for the inclusive
hard-photon production in association with a bottom- and
antibottom-quark pair at hadron colliders, $pp(p\pb) \rightarrow
b\bb\gam+X$ including the full effect of NLO QCD corrections as
described in Section~\ref{sec:nlo_2b_1b}.  We distinguish the case
where at least two $b$ jets are identified in the final state
($\gamma+2b+X$, or 2$b$-tag), and the case where at least one $b$ jet
is identified in the final state ($\gamma+b+X$, or 1$b$-tag). For the
1$b$-tag case, we compare the results obtained from our FFS/4FNS
calculation with the results obtained from the VFS/5FNS calculation at
NLO in QCD. Finally, we provide a first comparison of the
FFS/4FNS results for $\gamma+b+X$ with the measurements by the CDF
and D0 collaborations.

\subsection{The Setup}
\label{subsec:results_setup}
The numerical results for $b\bb\gamma$ production are presented for
proton-proton collisions at the LHC with $\sqrt{s}=8$ TeV and
proton-antiproton collisions at the Tevatron with $\sqrt{s}=1.96$ TeV.
The following SM parameters are used in the numerical evaluation,
\begin{center}
\begin{tabular}{cc}
\hline
\hline
Parameter & Value \\
\hline
$m_b$ & 4.62 GeV \\
$m_t$ & 173.2 GeV \\
$\alpha$ & 1/137 \\
\hline
\end{tabular}
\end{center}
where $m_t$, $m_b$, and $\alpha$ are the top-quark mass, bottom-quark
mass and electromagnetic coupling constant respectively.  The bottom
quark is treated as massive, with the number of light quarks entering
the fermion loop set to $\nlf=$ 4.  This means that any fermion loop
that enters in the virtual corrections consists of four light-quark,
one bottom-quark, and one top-quark loop.  The LO results use the
CTEQ6L1 PDF set \cite{Pumplin:2002vw} and the one-loop
evolution of the strong coupling, $\as$, with $\as^{\lo}(M_Z)=0.13$,
while the NLO results use the CT10nlo\_nf4 PDF set \cite{Lai:2010vv}
and the two-loop evolution of $\as$, with $\as^{\nlo}(M_Z,n_f=4)=0.1127$.
The renormalization and factorization scales are set equal to one
another, and the central scale is chosen to be a dynamical scale
given by the transverse momentum of the photon, i.e.
\begin{eqnarray}
\label{eq:dyn_scale}
&& \muR=\muF=\mu_0=p_T(\gam),
\end{eqnarray} 
for both the $2b$- and $1b$-tag case. We have explored other
possibilities and will comment on our choice in
Sec.~\ref{subsec:bbgam_2b}. The residual scale dependence of the LO
and NLO cross sections is studied by varying $\mu=\muR=\muF$ by a
factor of four around the central value given in
Eq.~(\ref{eq:dyn_scale}). Independent variation of $\muR$ and $\muF$
could be considered but the choice of a dynamical scale makes it
unpractical. We have instead allowed for a pretty conservative
variation by a factor of four around the central value $\mu_0$
(instead of the traditional factor of two). 

The selection cuts for the photon are: $p_T(\gam) > 30$~GeV and
$|\eta(\gam)|<1$ for the
Tevatron~\cite{Abazov:2012ea,Aaltonen:2013coa}, and $p_T(\gam) >
25$~GeV and $|\eta(\gam)|<$ 1.37 for the LHC~\cite{ATLAS:2012ar}. We notice that
the photon's rapidity ($y(\gam)$) and pseudorapidity ($\eta(\gam)$)
coincide. We also notice that the Tevatron D0 experiment also
considered a forward rapidity region, $1.5<|\eta(\gam)|<2.5$. We will
not introduce it as a default in presenting most of the results of our
study but we will consider it in Sec.~\ref{subsec:1bgam_vs_tevdata} to
compare with D0 results.  Both bottom-quark and the light-quark jets
are clustered using the anti-$k_T$ jet algorithm, with pseudo-cone
size $R=0.4$, and are required to pass the following selection cuts:
\begin{center}
\begin{tabular} { l l}
Tevatron: & $p_T(b,j) > 20$~GeV, $|\eta(b,j)| < 1.5$,
~\cite{Abazov:2012ea,Aaltonen:2013coa}\\ 
LHC: & $p_T(b,j) > 25$~GeV, $|\eta(b,j)| < 2.1$, ~\cite{Aad:2013vka}\\
\end{tabular}
\end{center}
where $\eta(b,j)$ denotes the pseudorapidity of the corresponding
jet. Since massless jets ($j$) in our calculation are always
single-parton objects, their pseudorapidity and rapidity also
coincide.  Finally, since we consider inclusive observables, we
include events with both 0 and 1 identified light-parton jet in our
calculation.

As explained in Sec.~\ref{subsec:photon_isolation}, we use both a
fixed-cone and a smooth-cone isolation criteria~\cite{Frixione:1998jh}
to reduce the hadronic activity around the hard photon and
minimize the contribution from photon fragmentation. We will compare
the two approaches when relevant and show the remaining results using
as default one of the two options as specified.  
We notice that, in
contrast to the top-quark case ($t\tb\gam$), here the bottom quarks
have to be included in the hadronic energy contribution, following the
prescription described in Sec.~\ref{subsec:photon_isolation}.

\subsection{$pp(p\pb)\rightarrow b\bb\gam + X $: 
at least two $b$ jets identified in the final state}
\label{subsec:bbgam_2b}

In this section we present results for $b\bb\gam$ production where at
least two $b$ jets are tagged in the final state. At NLO in QCD this
includes configurations in which the jets consist of just a single $b$
(or $\bb$) quark as well as configurations in which the $b$ or
$\bb$ jet include also a light parton. To identify a hard photon we
have used both a smooth-cone and a fixed-cone isolation criterion and
noticed minimal differences in the results. Therefore in this section
we limit ourselves to results that have been obtained using the
fixed-cone isolation criterion in Eq.~\ref{eq:photoiso}, with the following parameters:
\begin{center}
\begin{tabular} { l l}
Tevatron: & $R_0 = 0.4$, $E_T^{\mathrm{max}}=2$~GeV, ~\cite{Aaltonen:2013coa}\\ 
LHC: & $R_0 = 0.4$, $E_T^{\mathrm{max}}=5$~GeV, ~\cite{Aad:2013vka}.\\
\end{tabular}
\end{center}
We also use Set II of Bourhis, Fontannaz and Guillet~\cite{Bourhis:1997yu} for the photon
fragmentation functions.

\begin{table}[t!]
\begin{center}
\caption{Total cross section for $pp(p\pb) \rightarrow b\bb\gam + X$
  production with at least two $b$ jets tagged in the final state at
  the Tevatron ($\sqrt{s}=1.96$ TeV) and the LHC ($\sqrt{s}=8$ TeV),
  at LO and NLO, together with their $K$-factor.  The uncertainties
  are due to the dependence on the renormalization/factorization
  scale obtained by evaluating the cross section at
  $\muR=\muF=p_T(\gamma)/4$ for the upper value and at $\muR=\muF=4 p_T(\gamma)$
  for the lower value.  The integration errors are well below 1\% .}
\renewcommand{\arraystretch}{2}
\begin{tabular}{ |c|c|c|c| }
\hline \hline
Collider  & $\sigma_\lo$ [pb] & $\sigma_\nlo$ [pb] & $K$-factor \\ \hline
Tevatron at $\sqrt{s}=1.96$~TeV & $ 2.18 ^{+43\%}_{-104\%}$  
& $3.06 ^{+30 \%}_{-40 \%}$ & $1-1.8$ \\
LHC at $\sqrt{s}=8$~TeV & $ 29.4 ^{+38 \%}_{-73 \%}$ 
& $ 46.6 ^{+30\%}_{-60 \%}$ & $1.5-1.8$  \\[0.2cm]
\hline \hline
\end{tabular}
\label{tab:gam2b_xsec}
\end{center}
\end{table}

\begin{figure}
\begin{center}
\includegraphics[width=7.5cm]{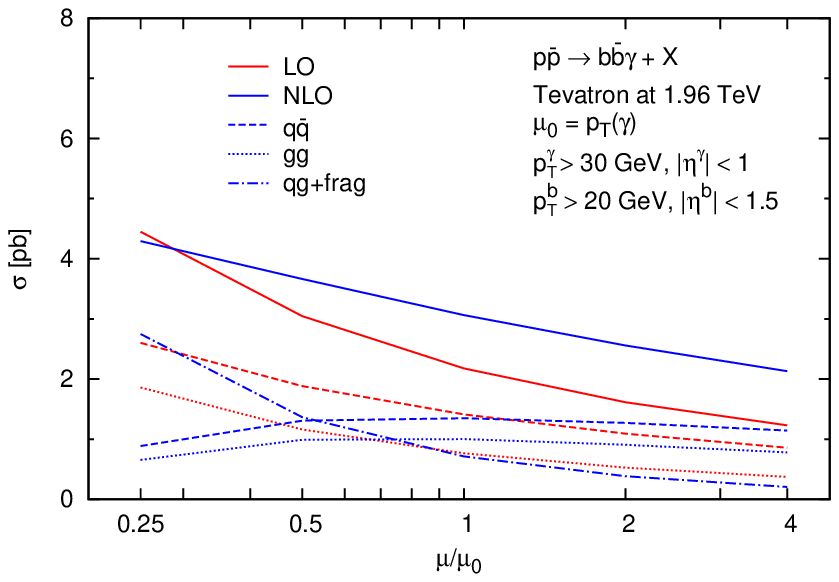}
\includegraphics[width=7.5cm]{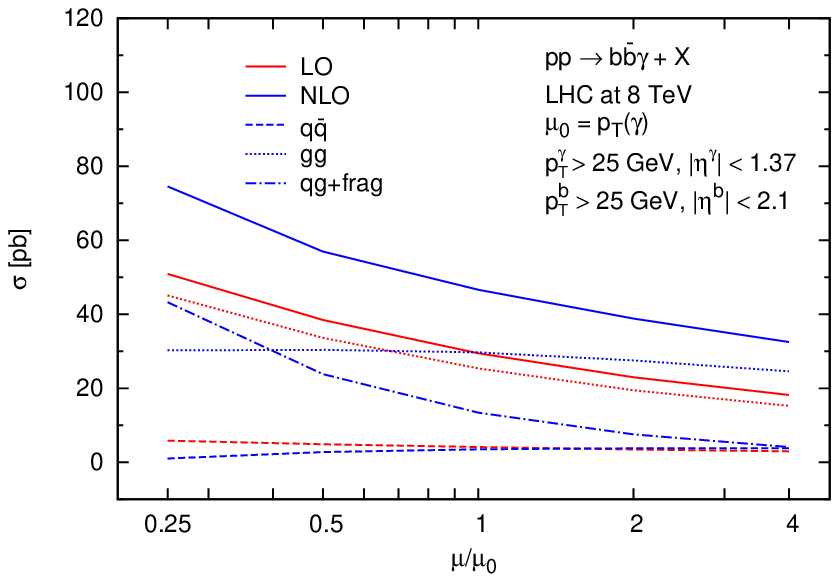}
\caption{Dependence of the LO (red) and NLO (blue) cross section for
  $pp \rightarrow b\bb\gam + X$ (at least two $b$ jets identified in
  the final state) on the renormalization/factorization scale at the
  Tevatron with $\sqrt{s}=1.96$~TeV (left) and at the LHC with
  $\sqrt{s}=8$~TeV (right). Both the total cross section (solid) and
  the contributions of the individual subprocesses, $q\qb$ (dashed),
  $gg$ (dotted), and $qg$ (dash-dotted) are shown.}
\label{fig:gam2b_mudep_sub}
\end{center}
\end{figure}

The impact of NLO QCD corrections on the perturbative stability of the
total cross section is well illustrated in Fig.~\ref{fig:gam2b_mudep_sub}
where we show the dependence of the total cross section on the
renormalization/factorization scale at the Tevatron and at the
LHC. The impact of QCD corrections depends on the chosen scale and is
always large. For renormalization/factorization scales that vary in
the $\mu_0/4 \leq \mu \leq 4\mu_0$ interval, we find $K$-factors
(defined as $K=\sigma_{NLO}/\sigma_{LO}$ with $\sigma_{NLO}$ and
$\sigma_{LO}$ calculated with the setup defined in this Section) that
range from $1$ to $1.8$ at the Tevatron, and from $1.5$ to $1.8$ at the
LHC. 

What is however more interesting is that the residual scale dependence
turns out to be quite substantial at both the Tevatron and LHC even
when NLO QCD corrections are included, with only a very little
improvement observed at the Tevatron.  To study the origin of the
strong scale dependence, we look at the scale dependence of the
different subprocesses that contribute to the total NLO cross section,
as shown in Fig~\ref{fig:gam2b_mudep_sub}.  At the Tevatron the
$q\qb$ subprocess dominates over the $gg$ subprocess, while the
opposite happens at the LHC, both at LO and at NLO. The $qg+\qb g$
subprocess (from now on denoted simply as $qg$) enters only at NLO
and, although never dominant, plays an important role at the LHC.  It
is interesting to notice that both at the Tevatron and at the LHC, the
scale dependence of the $q\qb$ and $gg$ subprocesses are improved at
NLO, as we can see from the plateau in the scale-dependence plot,
while the residual scale dependence is due to the $qg$ subprocess that
comes in at NLO as a tree level contribution and introduces therefore
a large scale dependence.

\begin{figure}
\begin{center}
\includegraphics[width=7.5cm]{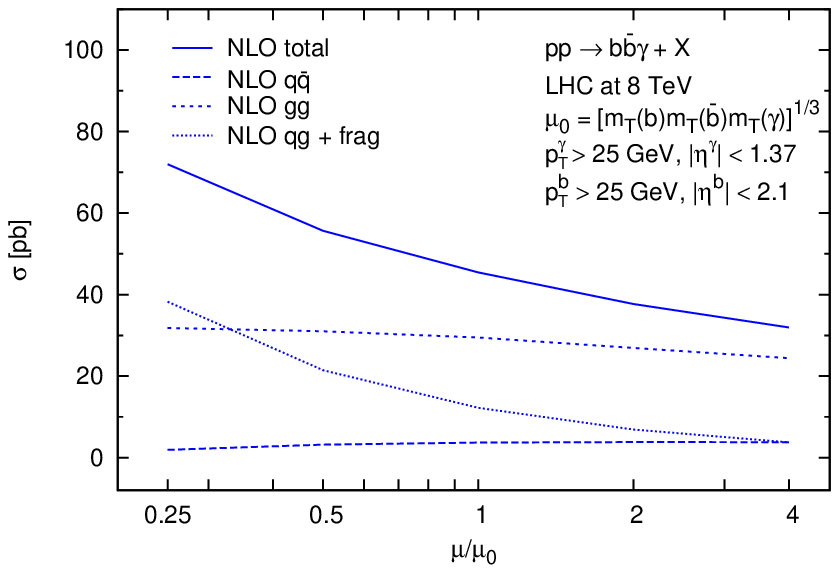} 
\includegraphics[width=7.5cm]{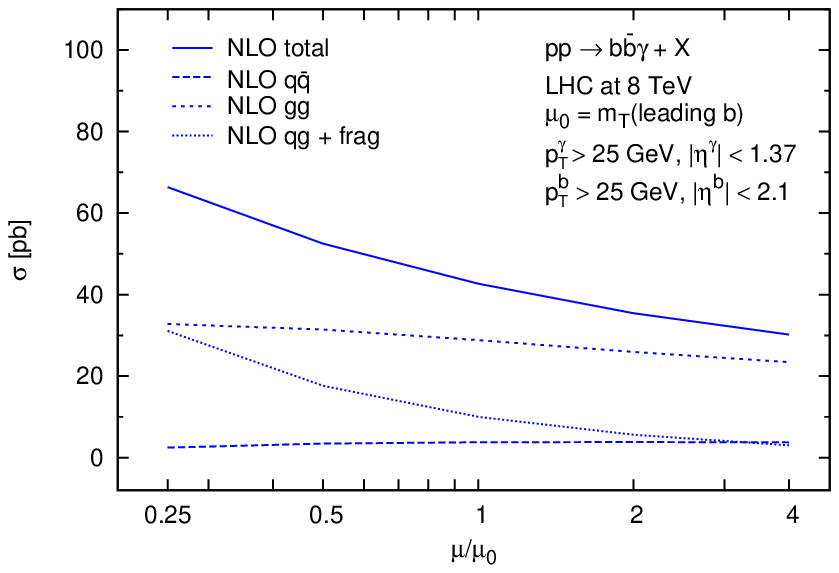}\\
\includegraphics[width=7.5cm]{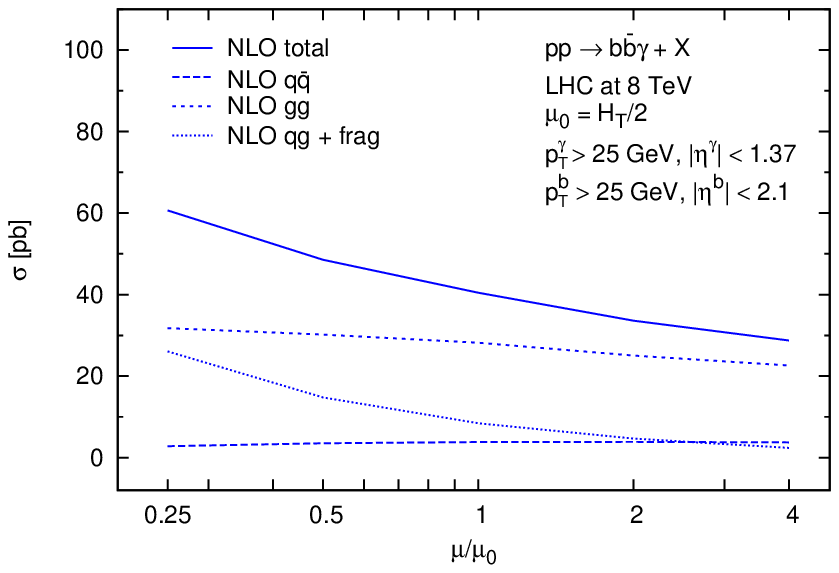}
\includegraphics[width=7.5cm]{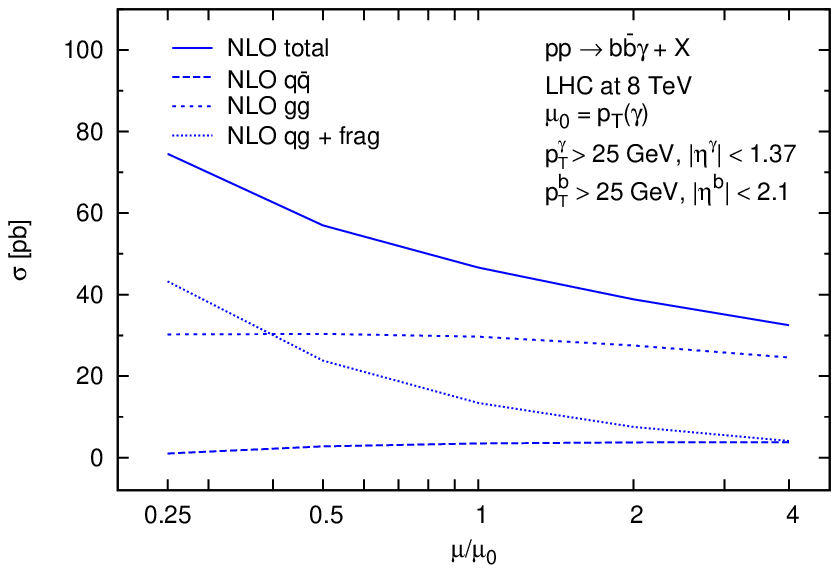}
\caption{Scale dependence of the NLO total cross section (solid) for
  $pp \rightarrow b\bb\gam+X$ (at least two $b$ jets identified in the
  final state) and of the individual $q\qb$- (dashed), $gg$- (dotted),
  and $qg$- (dash-dotted) subprocess contributions for four different
  choices of the central scale: $\mu_0 = (m_T(b) m_T(\bb)
  m_T(\gam))^{1/3}$ (top-left), $\mu_0 = m_T(\mathrm{leading}-b)$
  (top-right), $\mu_0 = H_T/2=\sum_{i=b,\bb,\gam} E_T^i/2$
  (bottom-left), $\mu_0 = p_T(\gam)$ (bottom-right), at the LHC with
  $\sqrt{s}=8$~TeV.}
\label{fig:gam2b_mudep_dyx}
\end{center}
\end{figure}

The choice of a dynamical scale is quite natural for this process,
since there is no preferred fixed hard scale to be chosen (like
$\mu=m_t$ for $t\bar{t}\gamma$ production). To investigate the
adequacy of choosing the photon transverse momentum, $p_T(\gam)$, as
dynamical scale, we have tried four different central-scale choices to
study the stability of each subprocess with respect to different
dynamical scales.  From the four plots that are shown in
Fig.~\ref{fig:gam2b_mudep_dyx}, where we take the LHC at
$\sqrt{s}=8$~TeV case as an example, it is evident that the NLO cross
sections are overall shifted when a different central scale is chosen.
By investigating the contribution from each subprocess, we see that
the shifting of the total NLO cross section is driven by the $qg$
subprocess and is therefore part of the theoretical uncertainty
introduced by the opening of this new channel at NLO. On the other
hand, both the $q\qb$ and $gg$ subprocesses are relatively insensitive
to the different choice of dynamical scale.  In presenting our results
for $b\bb\gam$ production we have therefore chosen the scale to be
fixed by $p_T(\gam)$ and have conservatively allowed it to vary by a
factor of four about the central value $\mu_0=p_T(\gam)$, as explained
earlier. Results are summarized in Table.~\ref{tab:gam2b_xsec}, where
we present both LO and NLO cross sections for the Tevatron at
$\sqrt{s}=1.96$~TeV and the LHC at $\sqrt{s}=8$~TeV.  The uncertainty
due to scale variation is obtained by evaluating the cross section at
$\muR=\muF=p_T(\gamma)/4$ ($\muR=\muF=4p_T(\gam)$) for the upper
(lower) value.


\begin{figure}
\begin{center}
\includegraphics[width=7.5cm]{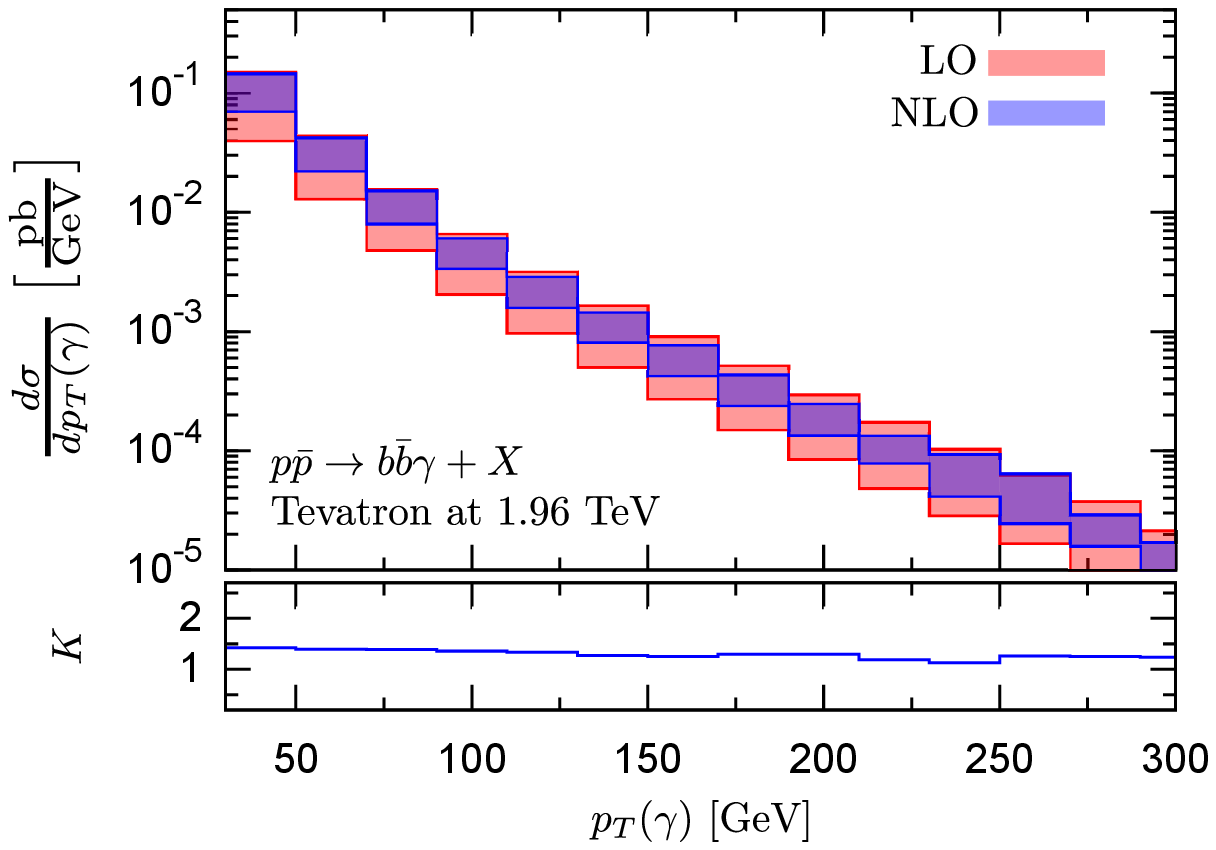}
\includegraphics[width=7.5cm]{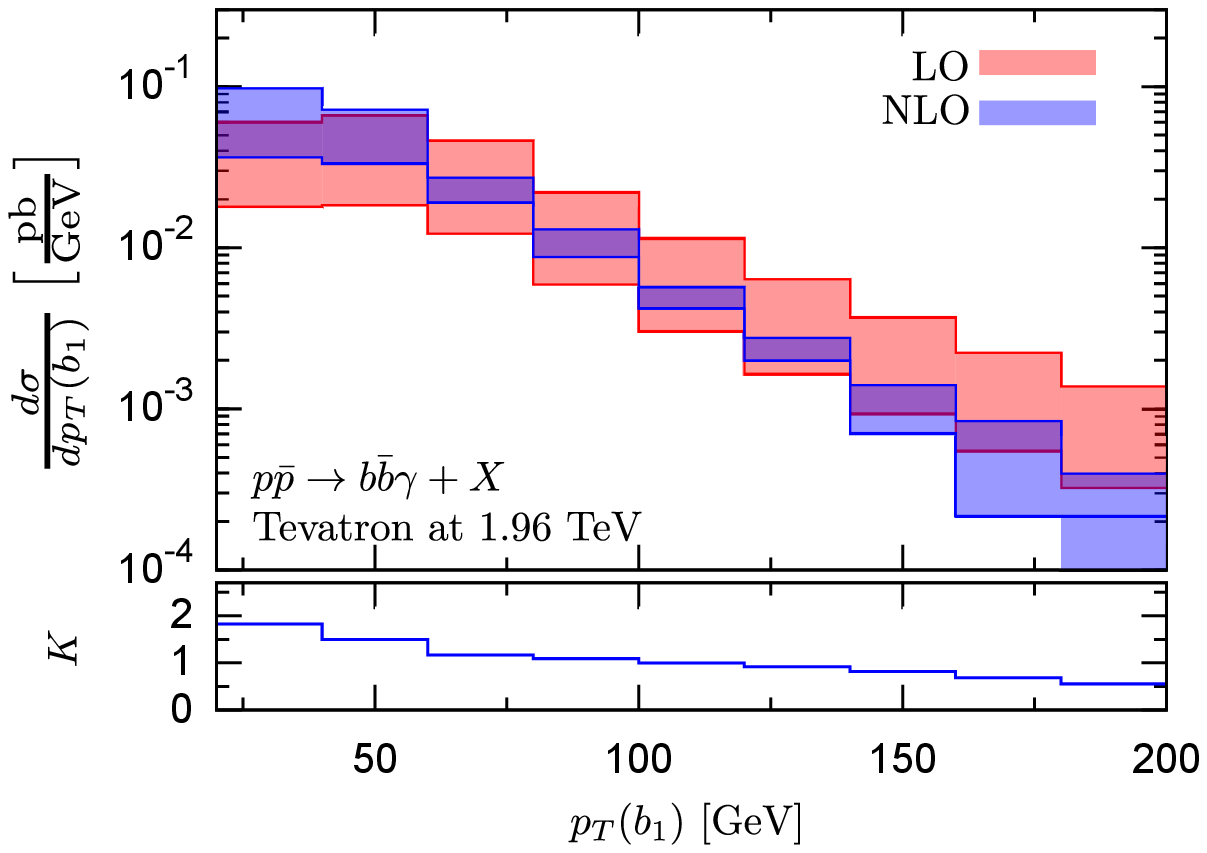}
\caption{The upper plots show the transverse-momentum distributions of
  the photon (left) and the leading $b$ jet (right) for $p\pb
  \rightarrow b\bb\gam+X$ (at least two $b$ jets identified in the
  final state) at the Tevatron with $\sqrt{s}=1.96$~TeV. The bands
  correspond to the variation of the renormalization and factorization
  scales in the interval $\mu_0/4 < \mu < 4\mu_0$.  The lower plots
  show the bin-by-bin $K$-factor for the corresponding distributions.}
\label{fig:gam2b_dist1_tev2}
\end{center}
\begin{center}
\includegraphics[width=7.5cm]{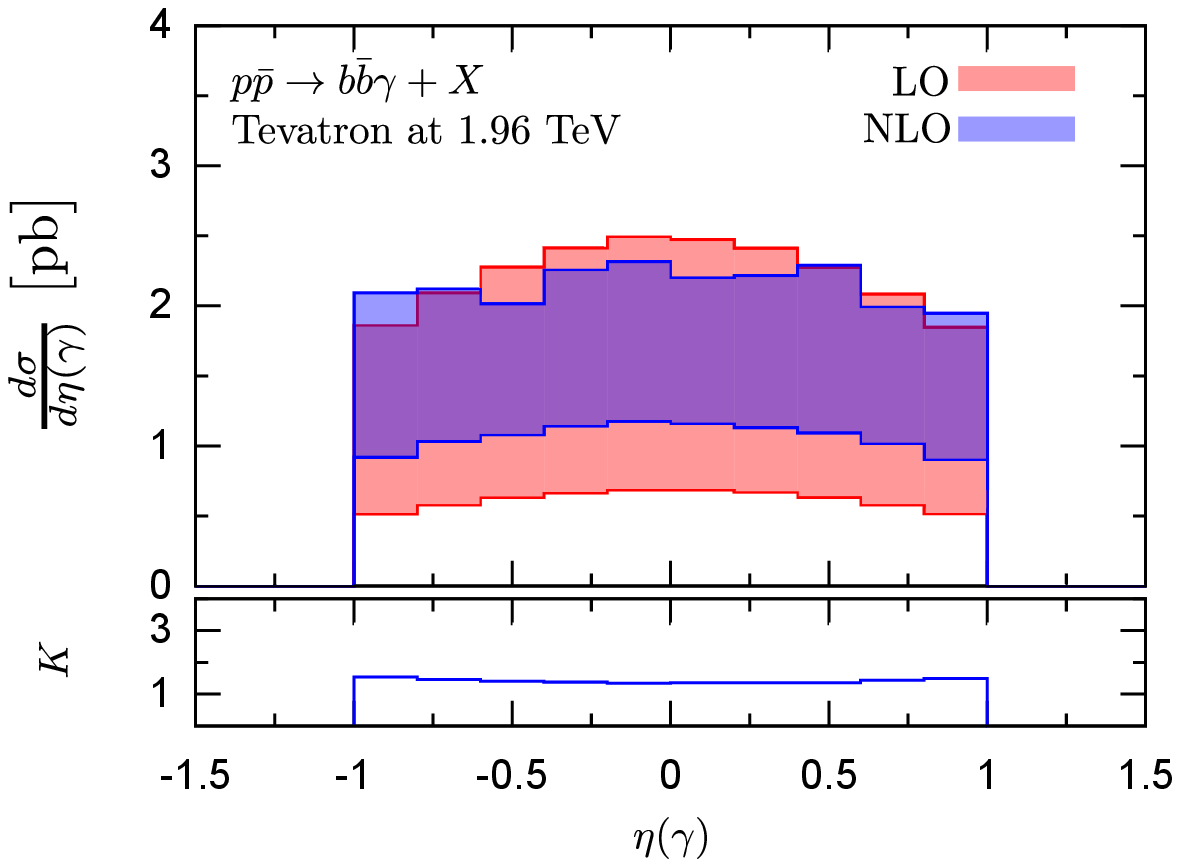}
\includegraphics[width=7.5cm]{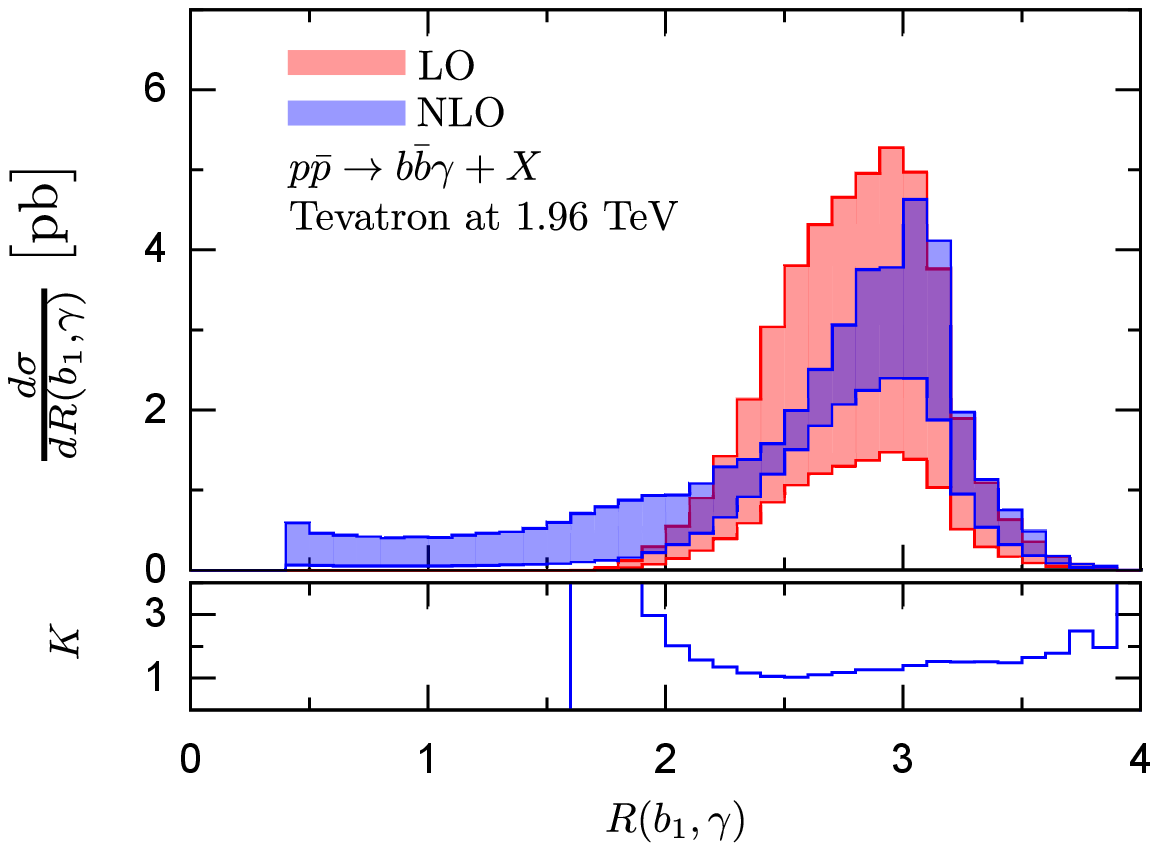}
\caption{The upper plots show the pseudorapidity distribution of
  final-state photon (left) and the separation between the leading $b$
  jet and the photon (right) for $p\pb \rightarrow b\bb\gam+X$ (at
  least two $b$ jets identified in the final state) at the
  Tevatron with $\sqrt{s}=1.96$~TeV. The bands correspond to the
  variation of the renormalization and factorization scales in the
  interval $\mu_0/4 < \mu < 4\mu_0$.  The lower plots show the
  bin-by-bin $K$-factor for the corresponding distributions.}
\label{fig:gam2b_dist2_tev2}
\end{center}
\end{figure}

\begin{figure}
\begin{center}
\includegraphics[width=7.5cm]{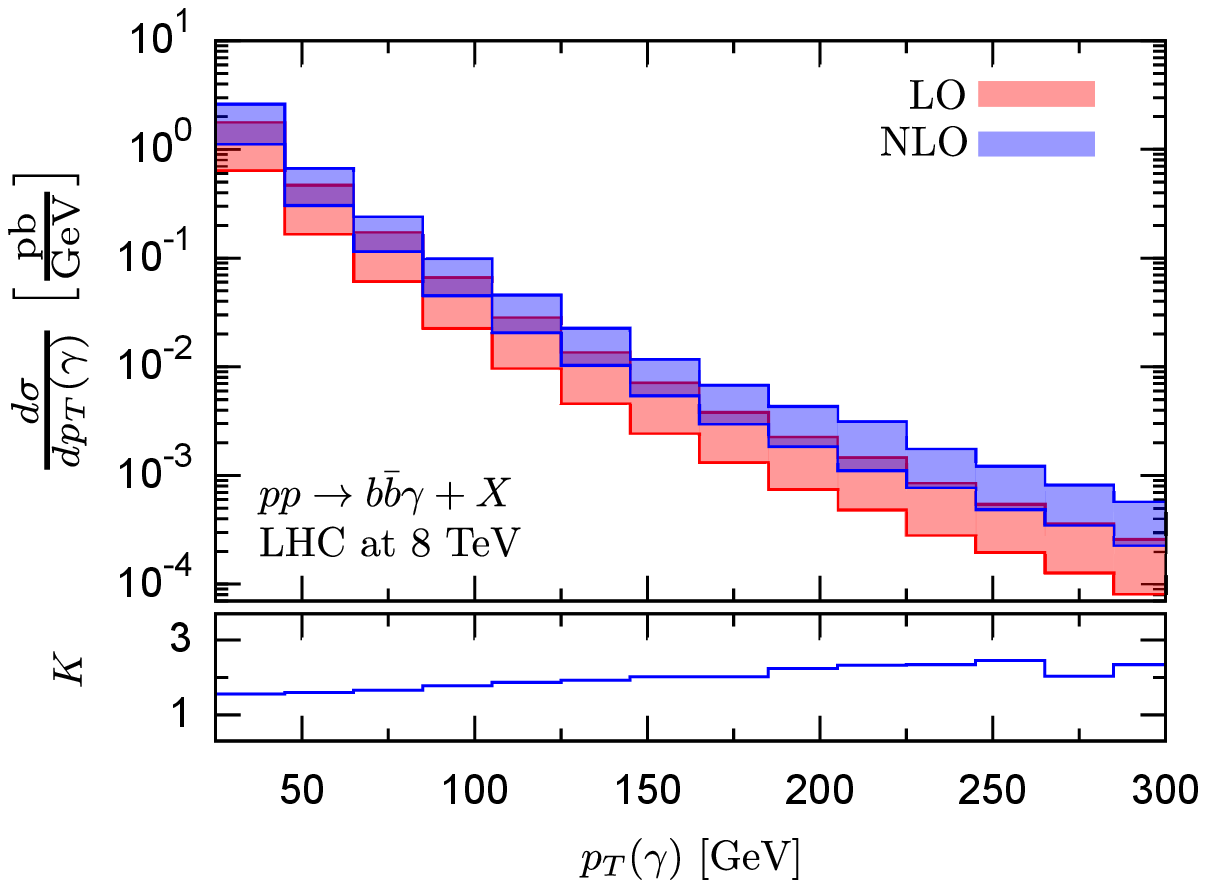} 
\includegraphics[width=7.5cm]{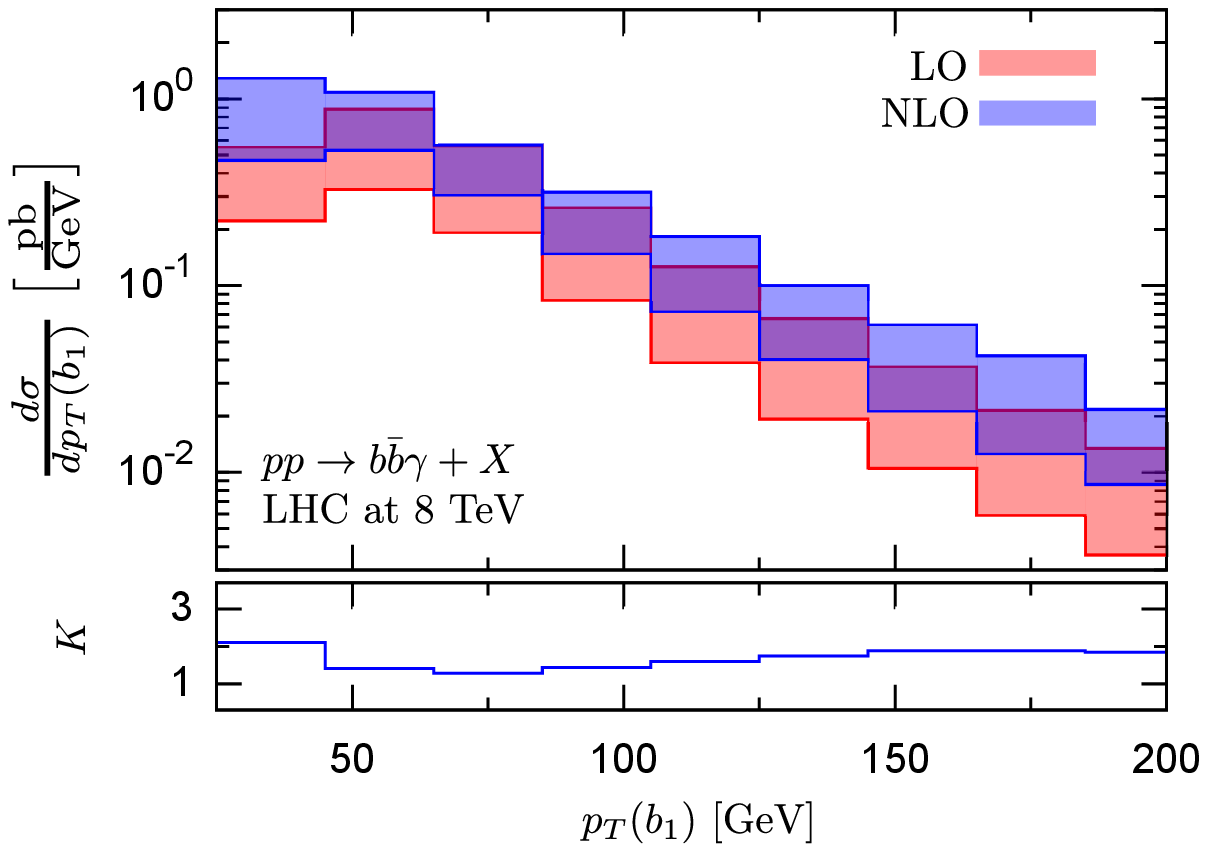}
\caption{The upper plots show the transverse-momentum distributions of
  the photon (left) and the leading $b$ jet (right) for $pp
  \rightarrow b\bb\gam+X$ (at least 2 $b$ jets identified in the
  final state) at the LHC with $\sqrt{s}=8$~TeV. The bands correspond
  to the variation of the renormalization and factorization scales in
  the interval $\mu_0/4 < \mu < 4 \mu_0$.  The lower plots show the
  bin-by-bin $K$-factor for the corresponding distributions.}
\label{fig:gam2b_dist1_lhc8}
\end{center}
\begin{center}
\includegraphics[width=7.5cm]{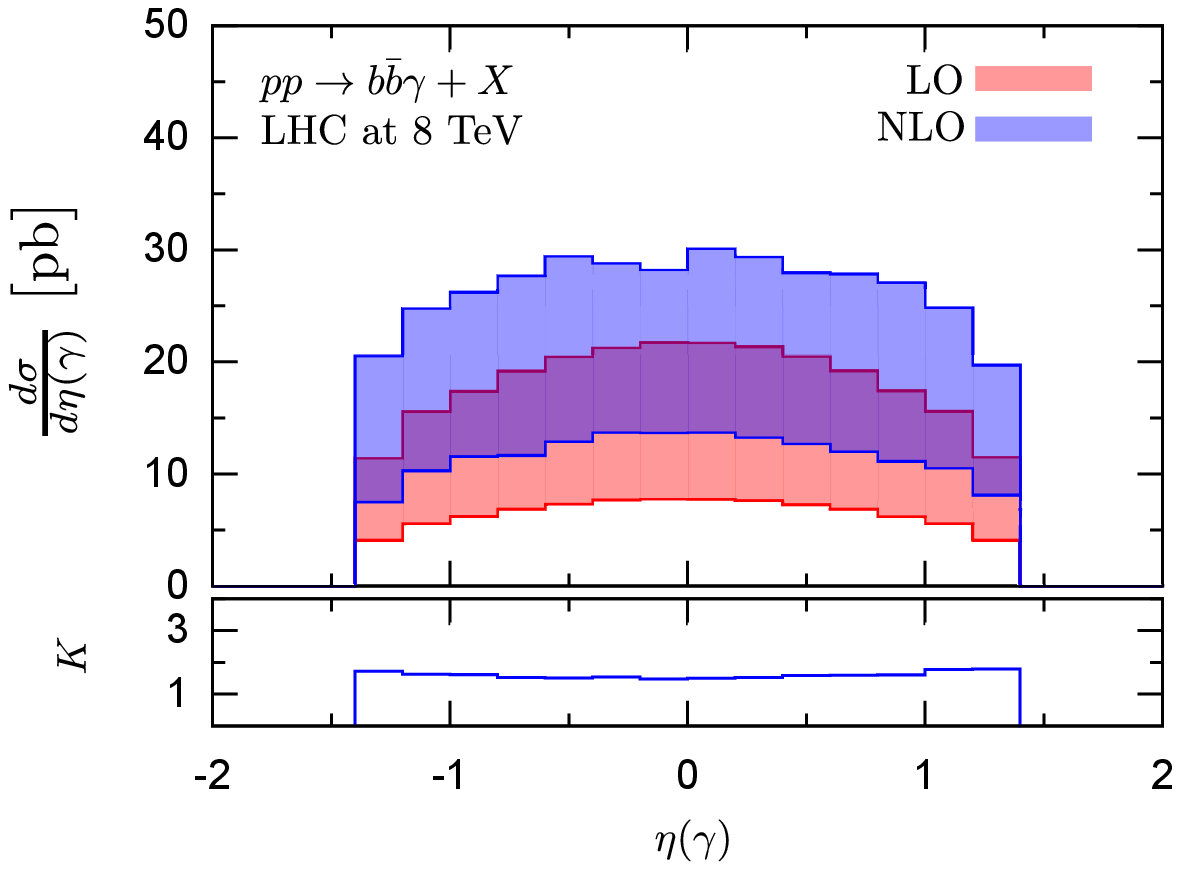} 
\includegraphics[width=7.5cm]{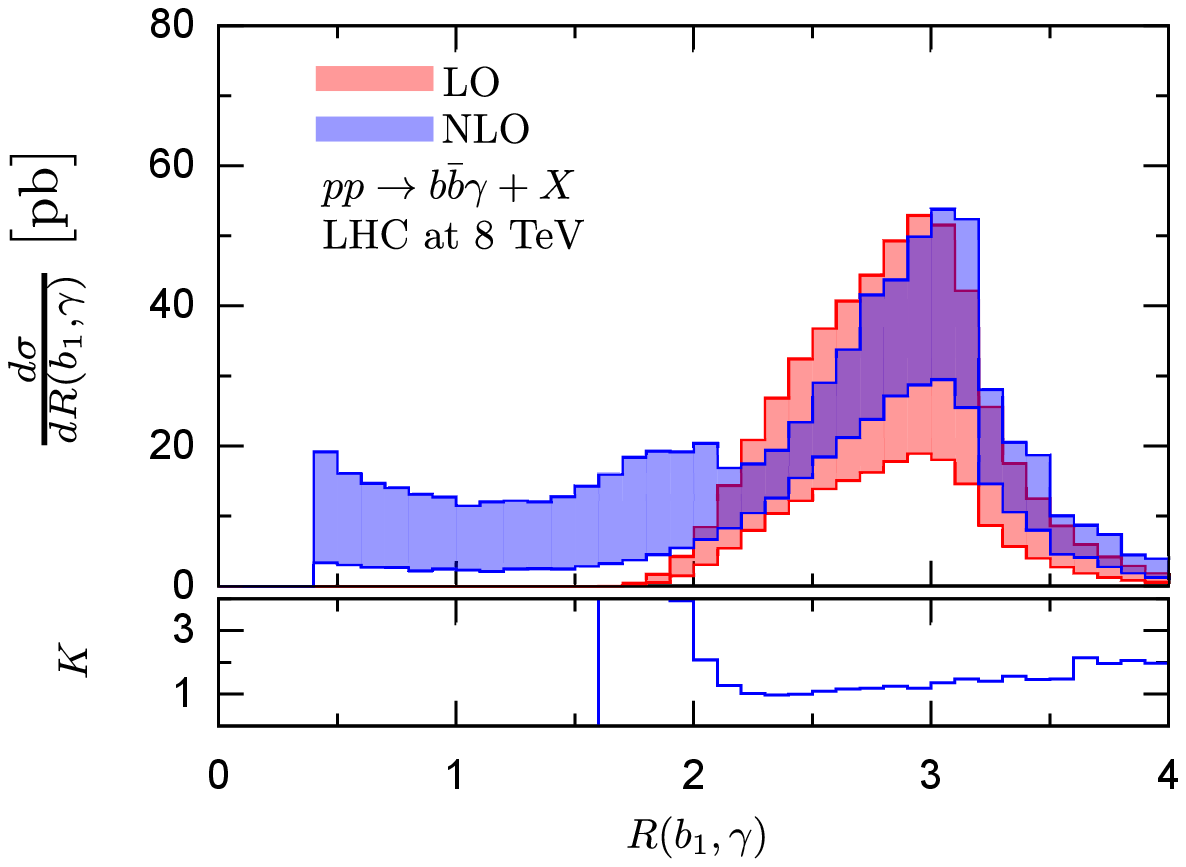}
\caption{The upper plots show the pseudorapidity distribution of the
  photon (left) and the separation between the leading $b$ jet and the
  photon (right) for $pp \rightarrow b\bb\gam+X$ (at least two $b$
  jets identified in the final state) at the LHC with
  $\sqrt{s}=8$~TeV. The bands correspond to the variation of the
  renormalization and factorization scales in the interval $\mu_0/4 <
  \mu < 4 \mu_0$.  The lower plots show the bin-by-bin $K$-factor for
  the corresponding distributions.}
\label{fig:gam2b_dist2_lhc8}
\end{center}
\end{figure}

We now turn to the phenomenologically more interesting case of
differential distributions.  In
Figs.~\ref{fig:gam2b_dist1_tev2}~and~\ref{fig:gam2b_dist2_tev2} and
Figs.~\ref{fig:gam2b_dist1_lhc8}~and~\ref{fig:gam2b_dist2_lhc8} we
show the photon and the leading $b$-jet ($b_1$) transverse-momentum
distributions as well as the photon pseudorapidity and photon to
leading $b$-jet separation ($R(\gamma,b_1)$) distributions at the
Tevatron and the LHC respectively. The bands in each figure represent
the variation of the differential cross section (bin-by-bin) when the
renormalization and factorization scales ($\muR=\muF$) are varied in
the range $\mu_0/4 \leq \mu \leq 4\mu_0$, with $\mu_0=p_T(\gam)$.  The
lower window of each figure gives the bin-by-bin $K$-factor.  The impact
of the NLO QCD corrections on the differential distributions is
sizable in both cases.  At the Tevatron, the $K$-factor for both the
$p_T(\gam)$ and $p_T(b_1)$ distributions decreases as $p_T$ grows,
while the opposite is true at the LHC.  For the photon pseudorapidity
distribution, the $K$-factor is quite large at the LHC and the shape
of the distribution at both the Tevatron and the LHC slightly changes
at NLO, becoming flatter due to less photon events that populate the
perpendicular direction with respect to the beam axis.  In the
$R(\gam,b_1)$ distribution we observe an accidental pinching of the
scale variation band at $R(\gamma,b_1) \sim 2.4-2.6$.  The $K$-factor
is also not well-defined for $R(\gam,b_1) < 1.5-1.7$, where at LO
there is no event.

\begin{figure}
\begin{center}
\includegraphics[width=7.5cm]{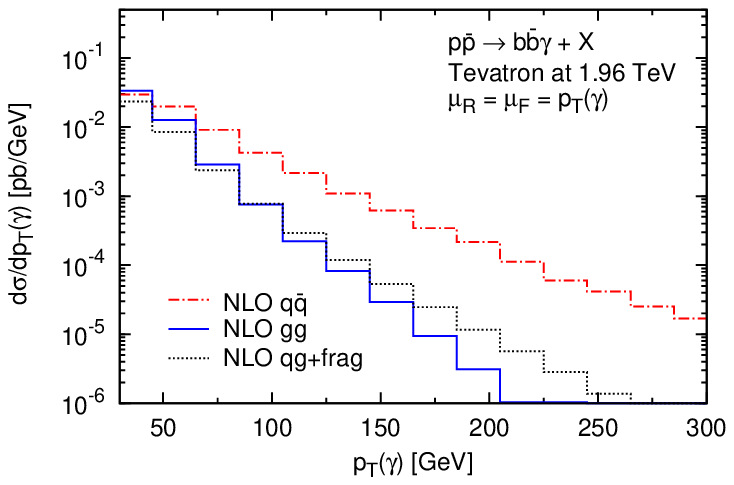}
\includegraphics[width=7.5cm]{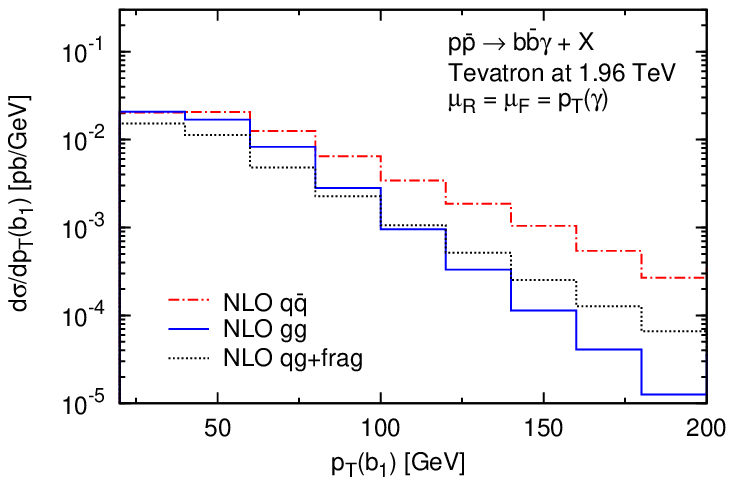}
\caption{Individual contributions of the $q\qb$ (red), $gg$ (blue), and
  $qg$ (black) subprocesses to the transverse-momentum distributions of
  the photon (left) and the leading $b$ jet (right) for $p\pb
  \rightarrow b\bb\gam+X$ (at least two $b$ jets identified in the
  final state) at the Tevatron with $\sqrt{s}=1.96$~TeV, and
  $\muR=\muF=p_T(\gam)$. }
\label{fig:gam2b_dist_tev2_sub}
\end{center}
\end{figure}

\begin{figure}
\begin{center}
\includegraphics[width=7.5cm]{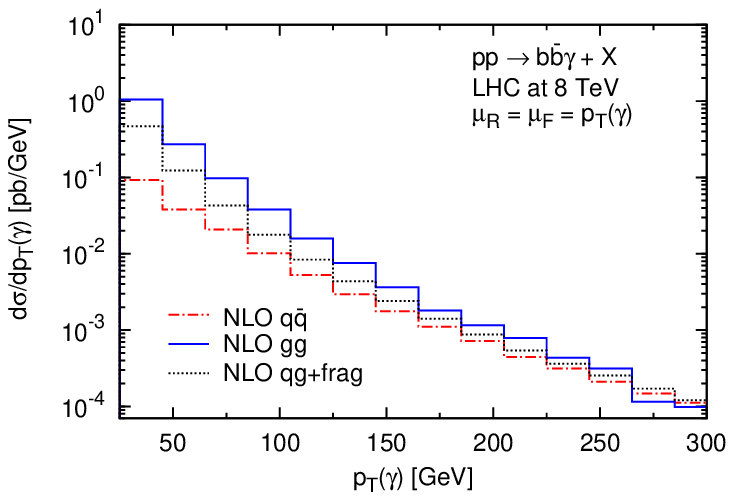} 
\includegraphics[width=7.5cm]{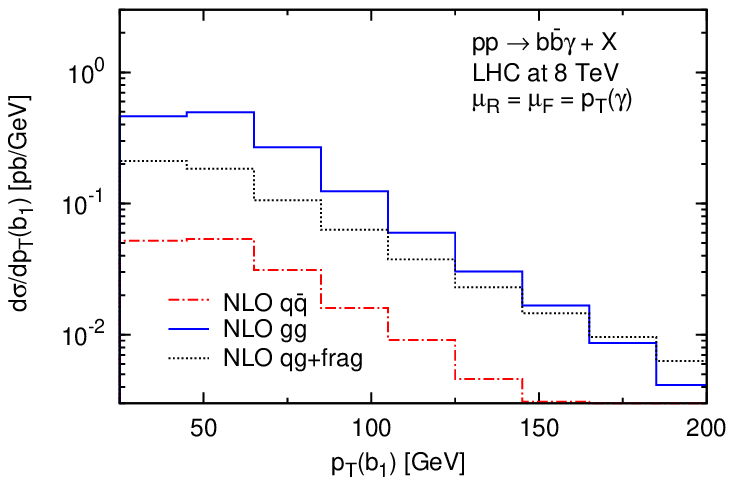}
\caption{Individual contributions of the $q\qb$ (red), $gg$ (blue), and
  $qg$ (black) subprocesses to the transverse-momentum distributions of
  the photon (left) and the leading $b$ jet (right) for $pp
  \rightarrow b\bb\gam+X$ (at least two $b$ jets identified in the
  final state) at the LHC with $\sqrt{s}=8$~TeV, and
  $\muR=\muF=p_T(\gam)$.}
\label{fig:gam2b_dist_lhc8_sub}
\end{center}
\end{figure}

The strong residual scale dependence at NLO is also manifest in the
differential distributions, in particular at the LHC where in
Fig.~\ref{fig:gam2b_dist1_lhc8}, both for $p_T(\gamma)$ and
$p_T(b_1)$, the NLO bands are as large as the LO bands, in the whole
$p_T$ region.  In contrast, at the Tevatron, the NLO bands are as
large as the LO bands at low $p_T$ and as the $p_T$ increases the NLO
bands are noticeably shrinking. This can be understood by looking at
the contribution of the different subprocesses to the NLO differential
distribution as shown in Figs.~\ref{fig:gam2b_dist_tev2_sub} and
\ref{fig:gam2b_dist_lhc8_sub}.  At the LHC, both for $p_T(\gam)$ and
$p_T(b_1)$, although the $gg$ subprocess dominates, the contribution
of the $qg$ subprocess, which suffers from strong scale dependence, is
quite large.  On the other hand, at the Tevatron, although in the
low-$p_T$ region the $gg$ and $qg$ subprocesses dominate, starting
from the intermediate-$p_T$ region, the $q\qb$ subprocess, which
receives quite considerable improvement in the scale dependence when
the NLO QCD corrections are included, starts to dominate while the
contributions of the $gg$ and $qg$ subprocesses drop rapidly.


\subsection{$pp(p\pb)\rightarrow b(\bb) \gam + X$: 
at least one $b$ jet identified in the final state}
\label{subsec:bbgam_1b}

\begin{table}[t!]
\begin{center}
\caption{Total cross section for $pp(p\pb) \rightarrow b(\bb)\gam + X$
  production with at least one $b$ jet tagged in the final state at
  the Tevatron ($\sqrt{s}=1.96$ TeV) and the LHC ($\sqrt{s}=8$ TeV),
  at LO and NLO, together with their $K$-factors.  The uncertainties
  are due to the dependence on the renormalization/factorization
  scale obtained by evaluating the cross section at
  $\mu=p_T(\gamma)/4$ for the upper value and at $\mu=4 p_T(\gamma)$
  for the lower value.  The integration errors are well below 1\%.}
\renewcommand{\arraystretch}{2}
\begin{tabular}{ |c|c|c|c| }
\hline \hline Collider & $\sigma_\lo$ [pb] & $\sigma_\nlo$ [pb] &
$K$-factor \\ \hline 
Tevatron at $\sqrt{s}=1.96$ TeV & $ 14.3 ^{+44 \%}_{-108\%}$ & 
$23.9 ^{+36 \%}_{-63 \%}$ & $1.3-1.9$ \\ 
LHC at$\sqrt{s}=8$ TeV & $ 299 ^{+35 \%}_{-59 \%}$ & $ 405 ^{+32 \%}_{-60 \%}$ & 
$1.3-1.4$ \\[0.2cm] \hline \hline
\end{tabular}
\label{tab:gam1b_xsec}
\end{center}
\end{table}

In this section we present the numerical results for $b\bb\gam$
production where at least one $b$ jet is tagged in the final state.
This set of events include events with 2 $b$ jets (see
Sec.~\ref{subsec:bbgam_2b}) as well as events with 1 $b$ jet that can
result from either a single $b$ or $\bb$ as well as from the
recombination of $b$ or $\bb$ with a light parton (quark or gluon), of
$b$ and $\bb$ themselves, or of $b$, $\bb$, and a light parton into a
$b$ jet that passes the selection cuts. As before, we have used both a
smooth-cone and a fixed-cone isolation criteria to identify a hard
photon. We notice a slightly bigger dependence on this choice with
respect to the $2b$-tagging case, but still very small. Therefore we
present most results using the fixed-cone isolation criterion
(see Eq.~\ref{eq:photoiso}), and
illustrate the comparison between the two choices in the case of the
$p_T(\gam)$ distribution.

\begin{figure}
\begin{center}
\includegraphics[width=7.5cm]{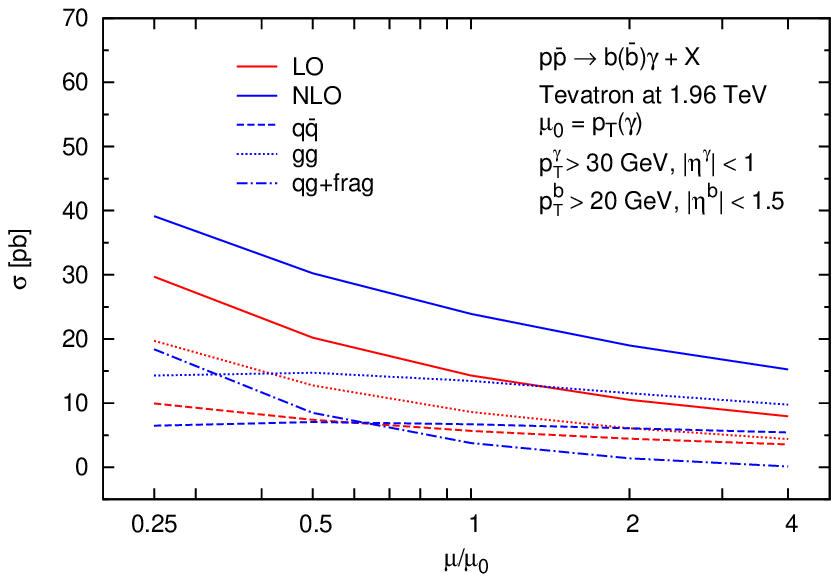}
\includegraphics[width=7.5cm]{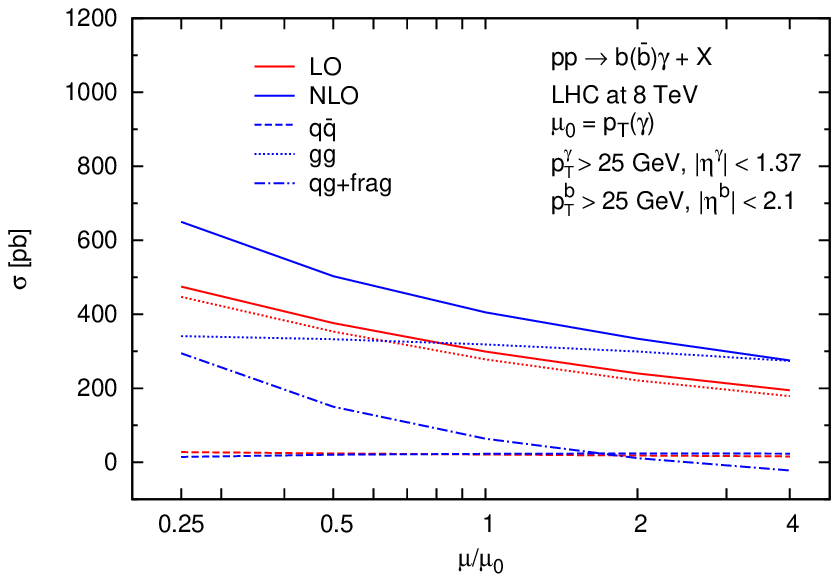}
\caption{Dependence of the LO (red) and NLO (blue) cross section for
  $pp \rightarrow b(\bb)\gam + X$ (at least one $b$ jet identified in
  the final state) on the renormalization/factorization scale at the
  Tevatron with $\sqrt{s}=1.96$~TeV (left) and at the LHC with
  $\sqrt{s}=8$~TeV (right). Both the total cross section (solid) and
  the contributions of the individual subprocesses, $q\qb$ (dashed),
  $gg$ (dotted), and $qg$ (dash-dotted) are shown.}
\label{fig:gam1b_mudep_sub}
\end{center}
\end{figure}

The scale dependence of the total cross section for the Tevatron and
the LHC is shown in Fig.~\ref{fig:gam1b_mudep_sub}, where the
contributions of the individual subprocesses are also given.  Similar
to the 2$b$-tag case, also in the 1$b$-tag case the impact of NLO QCD
corrections is substantial and the corresponding $K$-factors range
from $1.3$ to $1.9$ at the Tevatron and from $1.3$ to $1.4$ at the
LHC, when both renormalization and factorization scales are varied in
the $\mu_0/4 < \muR=\muF < 4\mu_0$ interval (for $\mu_0=p_T(\gam)$).
The improvement of the scale dependence is also not significant when
we go from LO to NLO and the residual scale dependence at NLO is also
due to the $qg$ channel, while the scale dependence of the $q\qb$ and
$gg$ channels are greatly improved when the NLO QCD corrections are
included, as shown in Fig.~\ref{fig:gam1b_mudep_sub}. 

\begin{figure}
\begin{center}
\includegraphics[width=7.5cm]{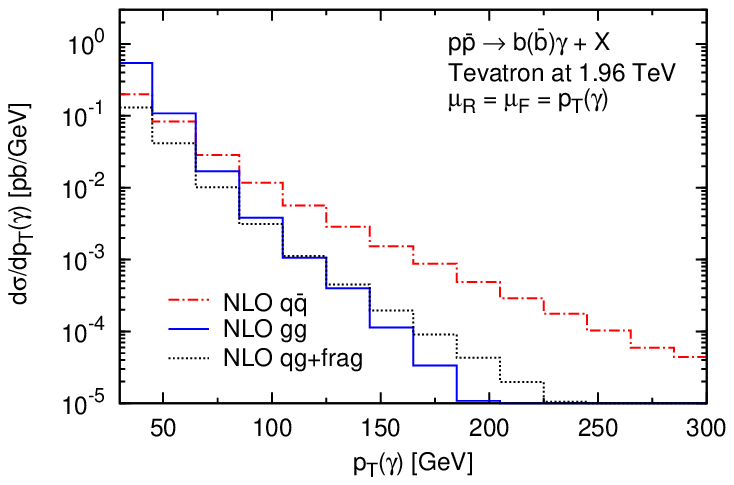}
\includegraphics[width=7.5cm]{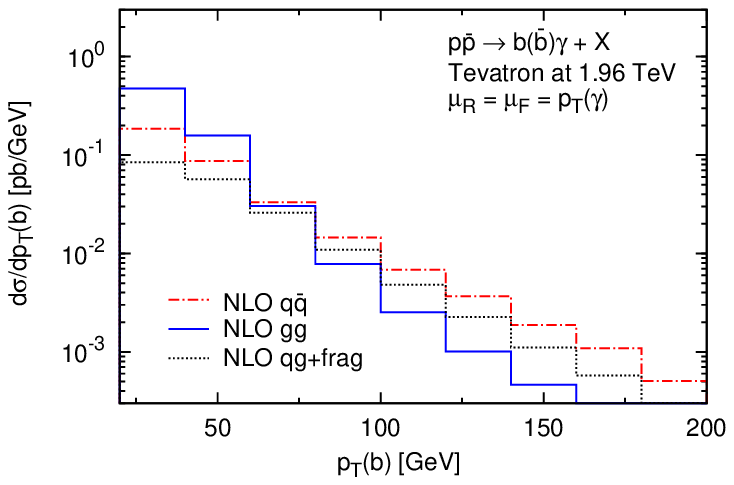}
\caption{Individual contributions of the $q\qb$ (red), $gg$ (blue) and
  $qg$ (black) subprocesses to the transverse-momentum distributions of
  the photon (left) and the $b$ jet (right) for $pp \rightarrow
  b(\bb)\gam+X$ (at least one $b$ jet identified in the final
  state) at the Tevatron with $\sqrt{s}=1.96$~TeV, and
  $\muR=\muF=p_T(\gam)$.}
\label{fig:gam1b_dist_tev2_sub}
\end{center}
\begin{center}
\includegraphics[width=7.5cm]{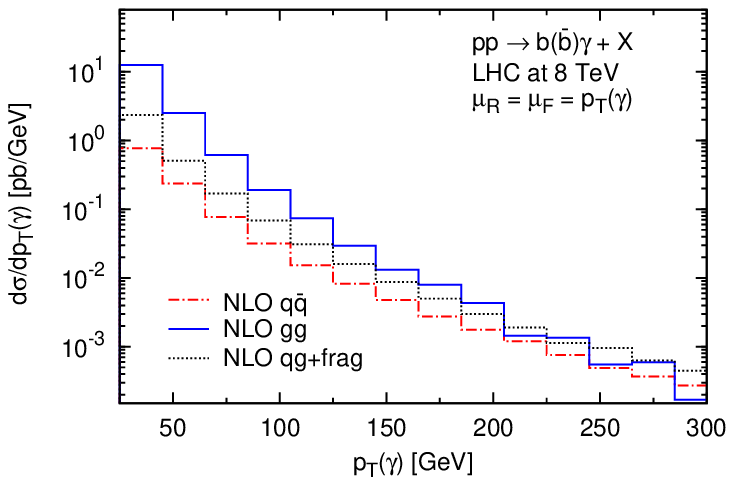} 
\includegraphics[width=7.5cm]{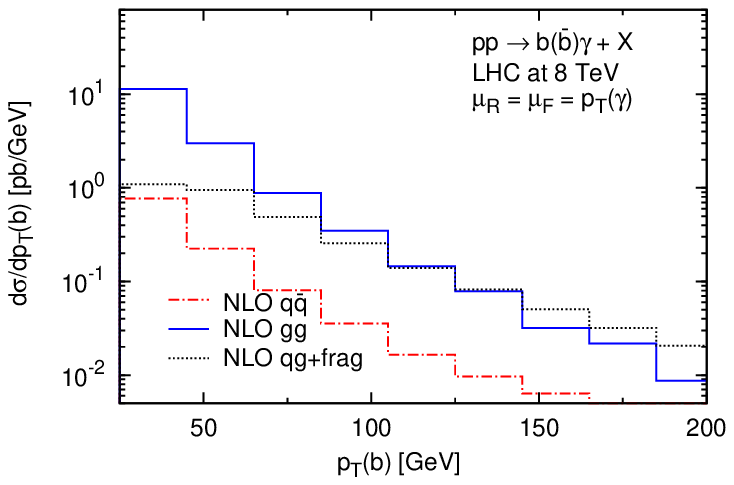}
\caption{Individual contributions of the $q\qb$ (red), $gg$ (blue) and
  $qg$ (black) subprocesses to the transverse-momentum distributions of
  the photon (left) and the $b$ jet (right) for $pp \rightarrow
  b(\bb)\gam+X$ (at least one $b$ jet identified in the final
  state) at the LHC with $\sqrt{s}=8$~TeV, and $\muR=\muF=p_T(\gam)$. }
\label{fig:gam1b_dist_lhc8_sub}
\end{center}
\end{figure}

We notice that the leading subprocesses in the total cross section are
now $gg$ and $qg$ at both at the Tevatron and the LHC. However, as
we will see in Fig.~\ref{fig:gam1b_dist_tev2_sub}, this is true at
the level of distributions only for the LHC, while at the Tevatron the
$q\qb$ subprocesses still dominates at medium and large $p_T(\gam)$ and
$p_T(b)$. This will be important to understand the comparison between
FFS/4FNS and VFS/5FNS that we will discuss in
Sec.~\ref{subsec:bbgam_4v5}.

\begin{figure}
\begin{center}
\includegraphics[width=7.5cm]{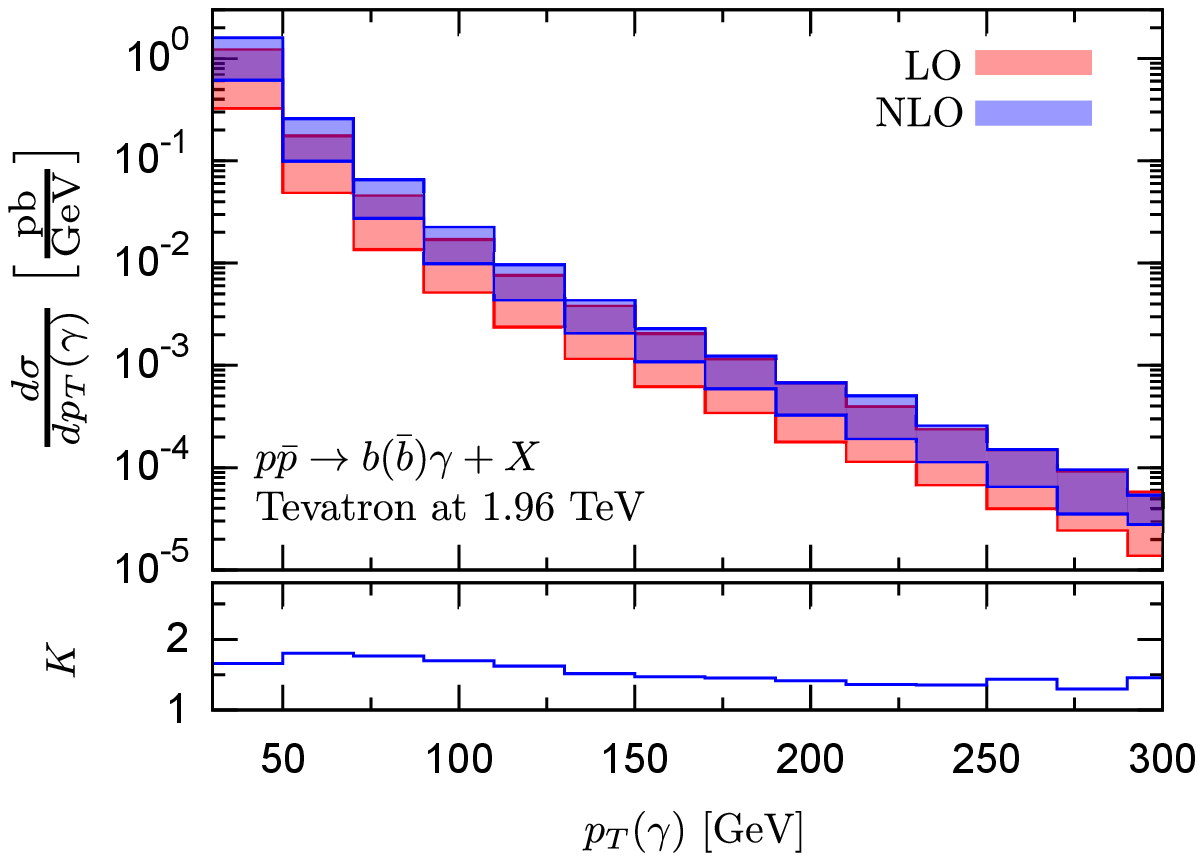}
\includegraphics[width=7.5cm]{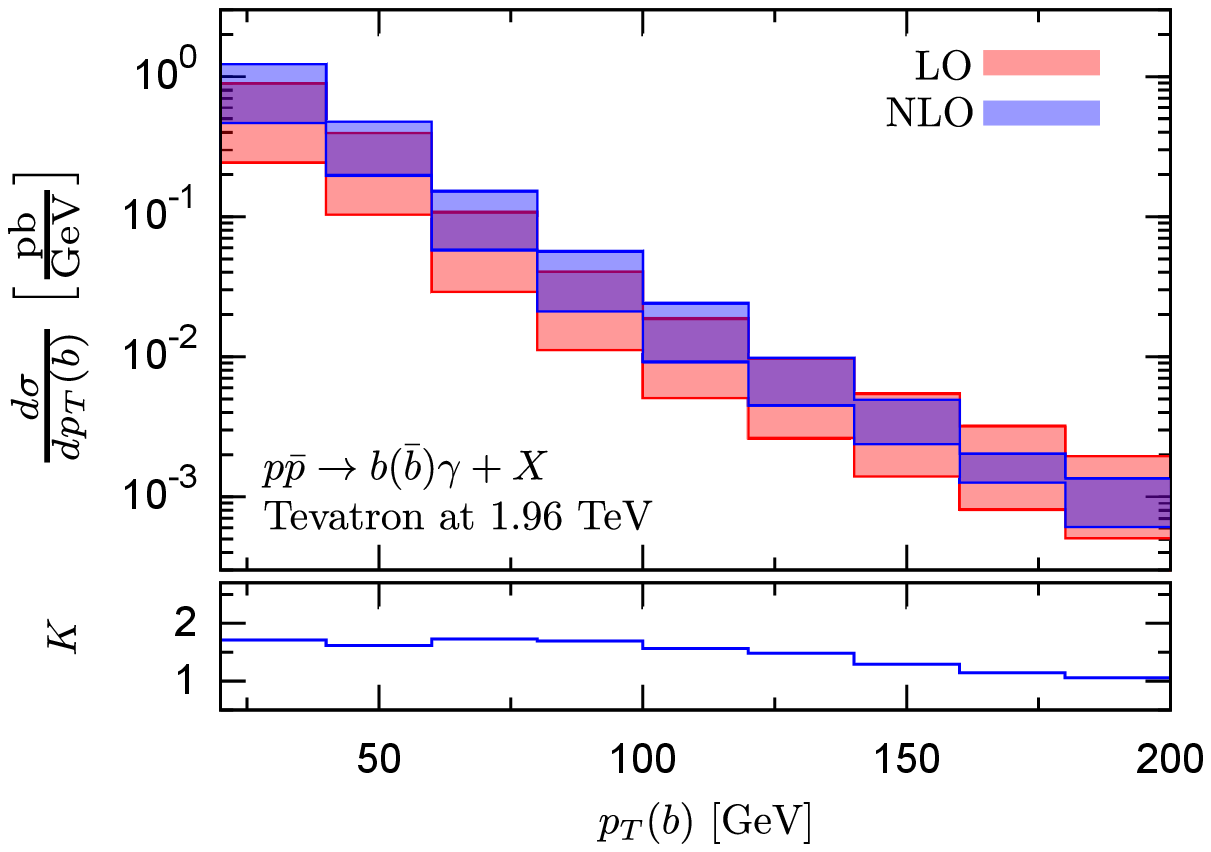}
\caption{The upper plots show the transverse-momentum distributions of
  the photon (left) and the $b$ jet (right) for $p\pb \rightarrow
  b(\bb)\gam+X$ (at least one $b$ jet identified in the final
  state) at the Tevatron with $\sqrt{s}=1.96$~TeV. The bands
  correspond to the variation of the renormalization and factorization
  scales in the interval $\mu_0/4 < \mu < 4 \mu_0$.  The lower plots
  show the bin-by-bin $K$-factor for the corresponding
  distributions. }
\label{fig:gam1b_dist1_tev2}
\end{center}
\begin{center}
\includegraphics[width=7.5cm]{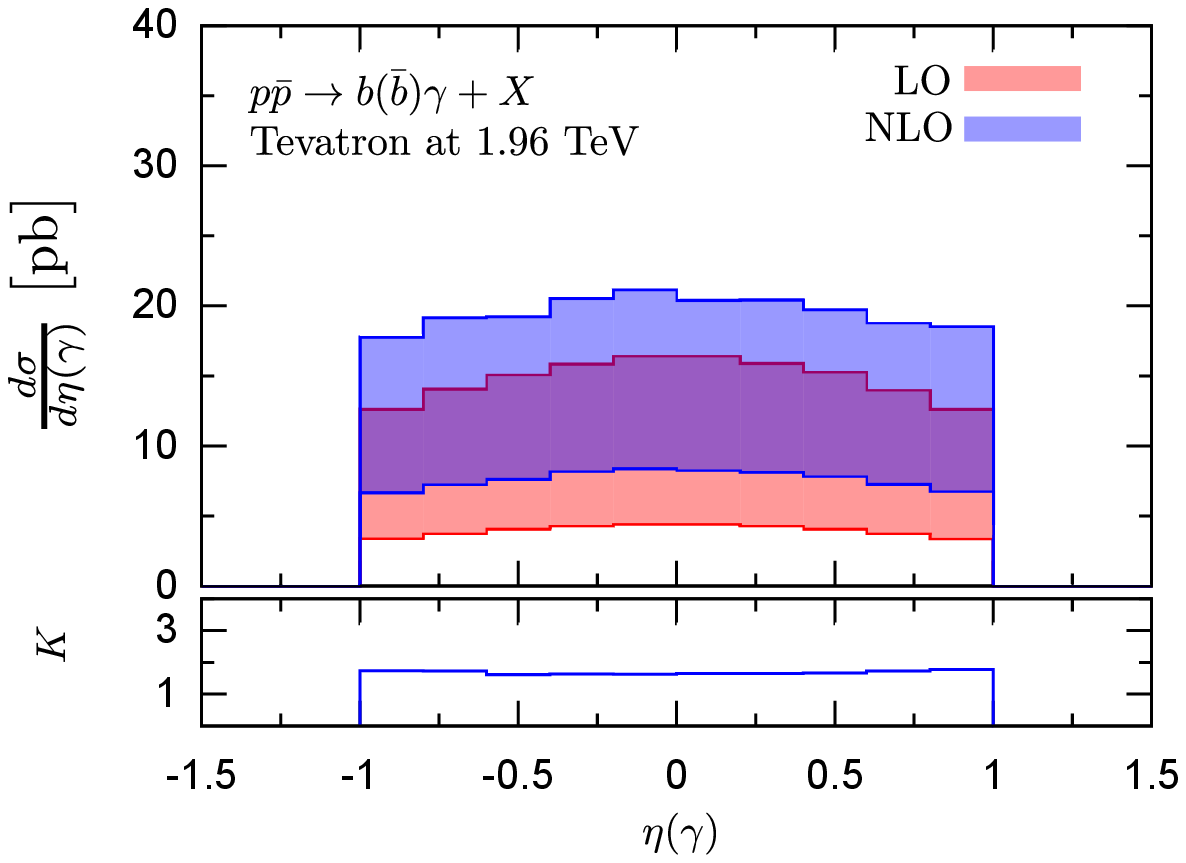}
\includegraphics[width=7.5cm]{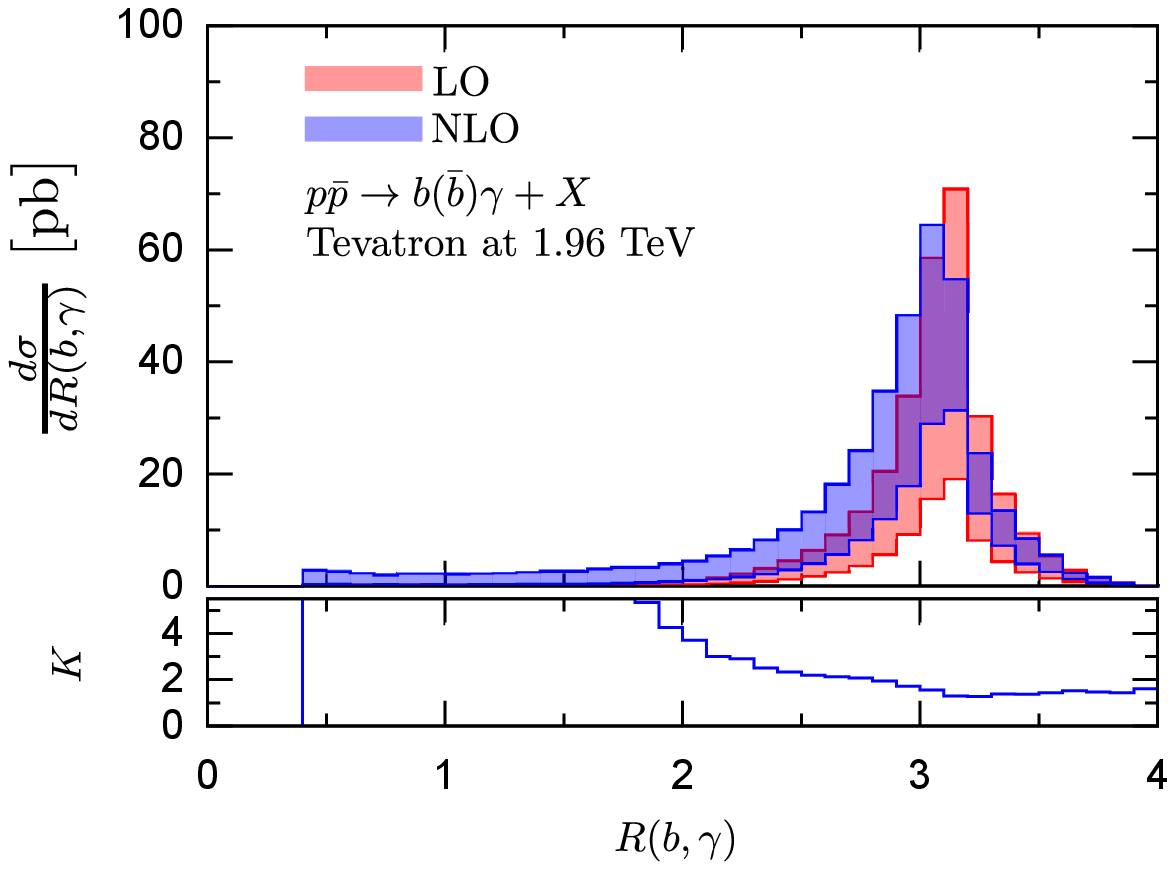}
\caption{The upper plots show the pseudorapidity distribution of the
  photon (left) and the separation between the $b$ jet and the photon
  (right) for $p\pb \rightarrow b(\bb)\gam+X$ (at least one $b$ jet
  identified in the final state) at the Tevatron with
  $\sqrt{s}=1.96$~TeV. The bands correspond to the variation of the
  renormalization and factorization scales in the interval $\mu_0/4 <
  \mu < 4 \mu_0$.  The lower plots show the bin-by-bin $K$-factor for
  the corresponding distributions.}
\label{fig:gam1b_dist2_tev2}
\end{center}
\end{figure}
\begin{figure}
\begin{center}
\includegraphics[width=7.5cm]{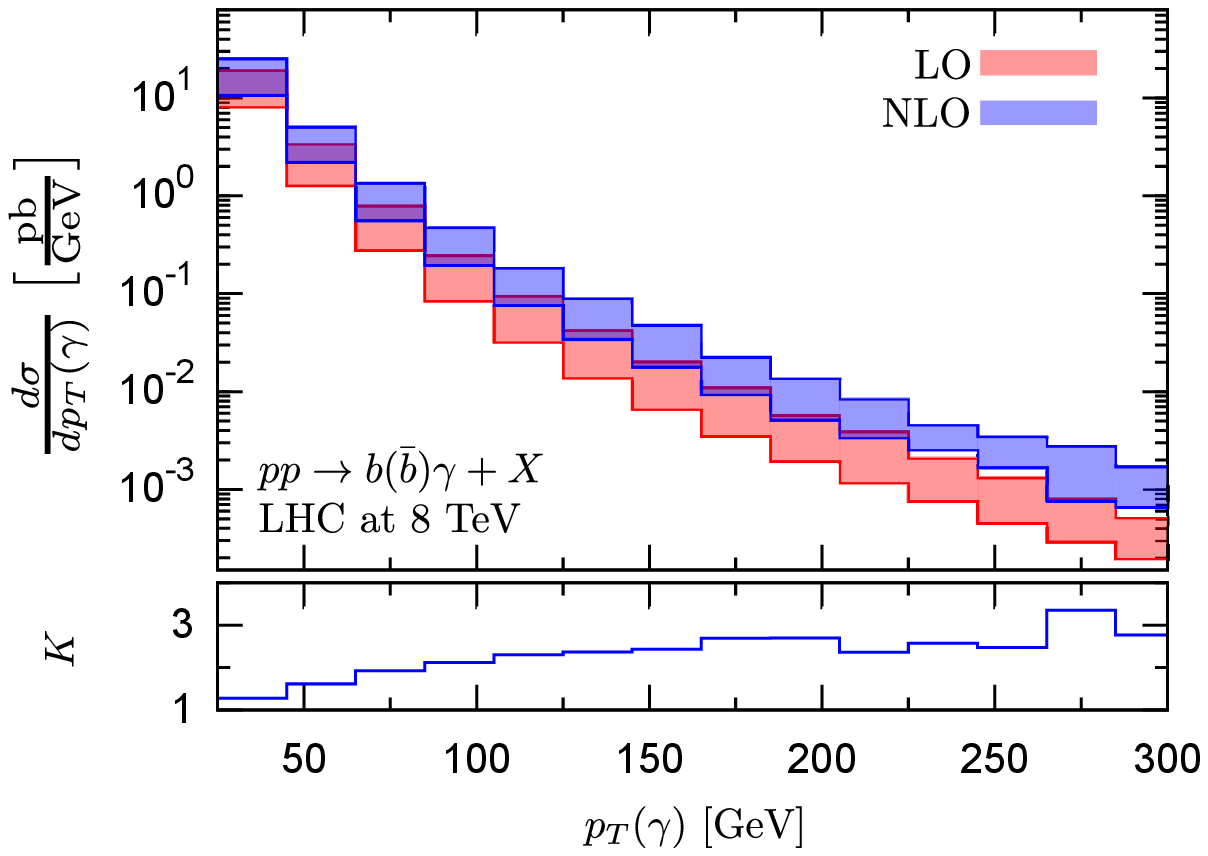} 
\includegraphics[width=7.5cm]{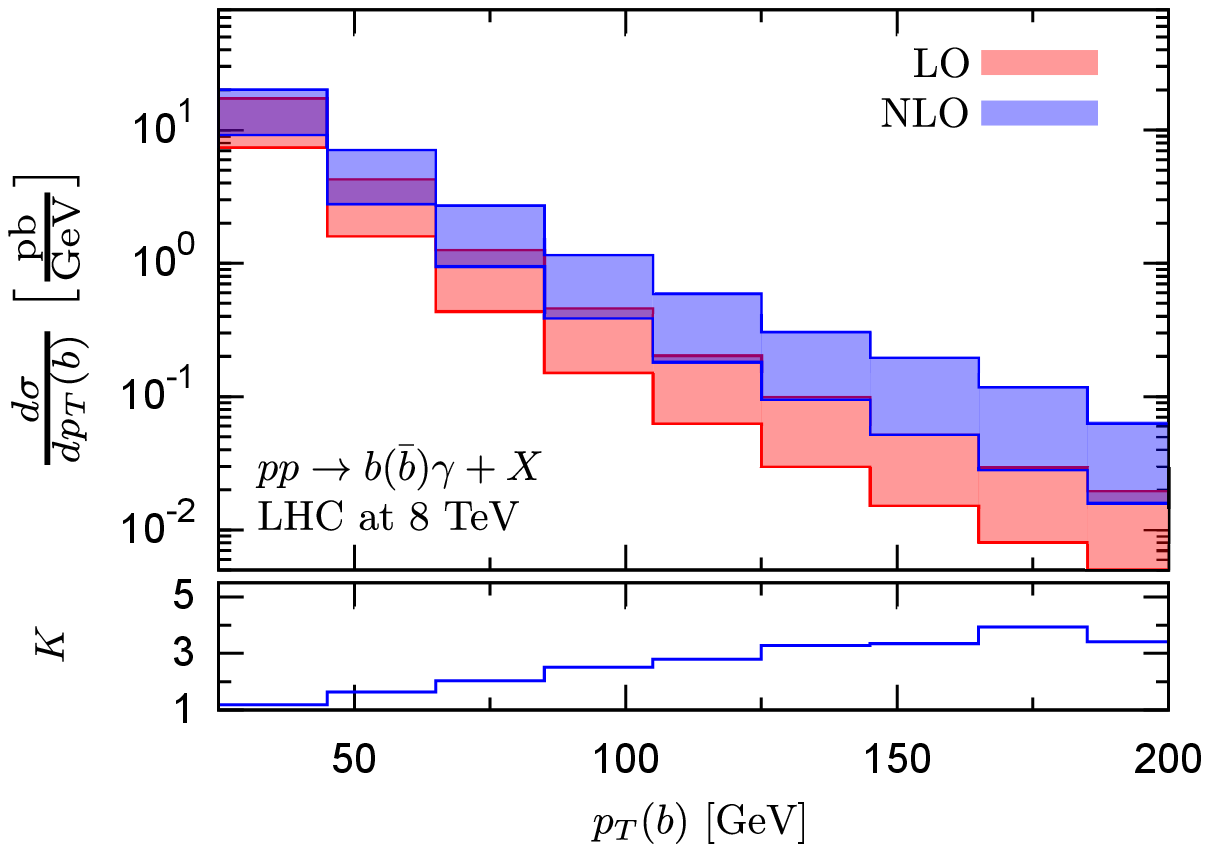}
\caption{The upper plots show the transverse-momentum distributions of
  the photon (left) and the $b$ jet (right) for $pp \rightarrow
  b(\bb)\gam+X$ (at least one $b$ jet identified in the final
  state) at the LHC with $\sqrt{s}=8$~TeV. The bands correspond to the
  variation of the renormalization and factorization scales in the
  interval $\mu_0/4 < \mu < 4 \mu_0$.  The lower plots show the
  bin-by-bin $K$-factor for the corresponding distributions.}
\label{fig:gam1b_dist1_lhc8}
\end{center}
\begin{center}
\includegraphics[width=7.5cm]{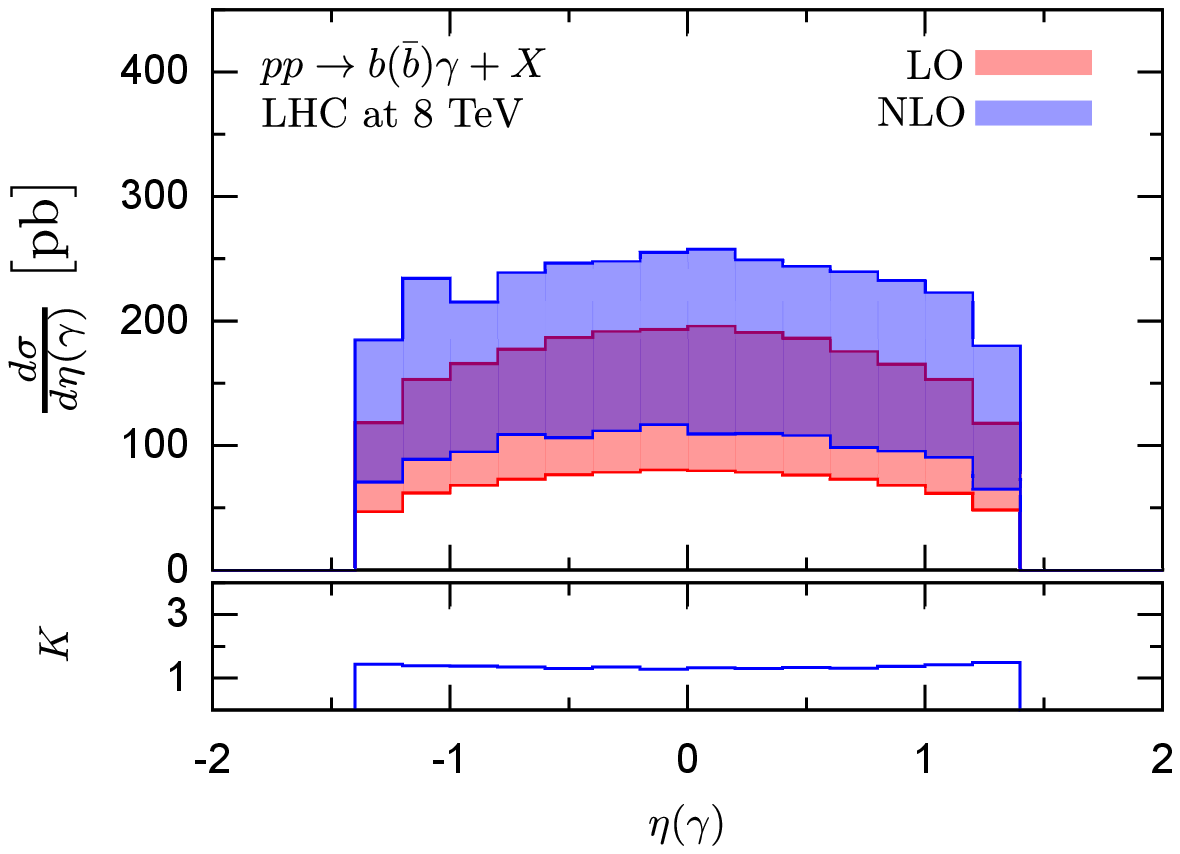} 
\includegraphics[width=7.5cm]{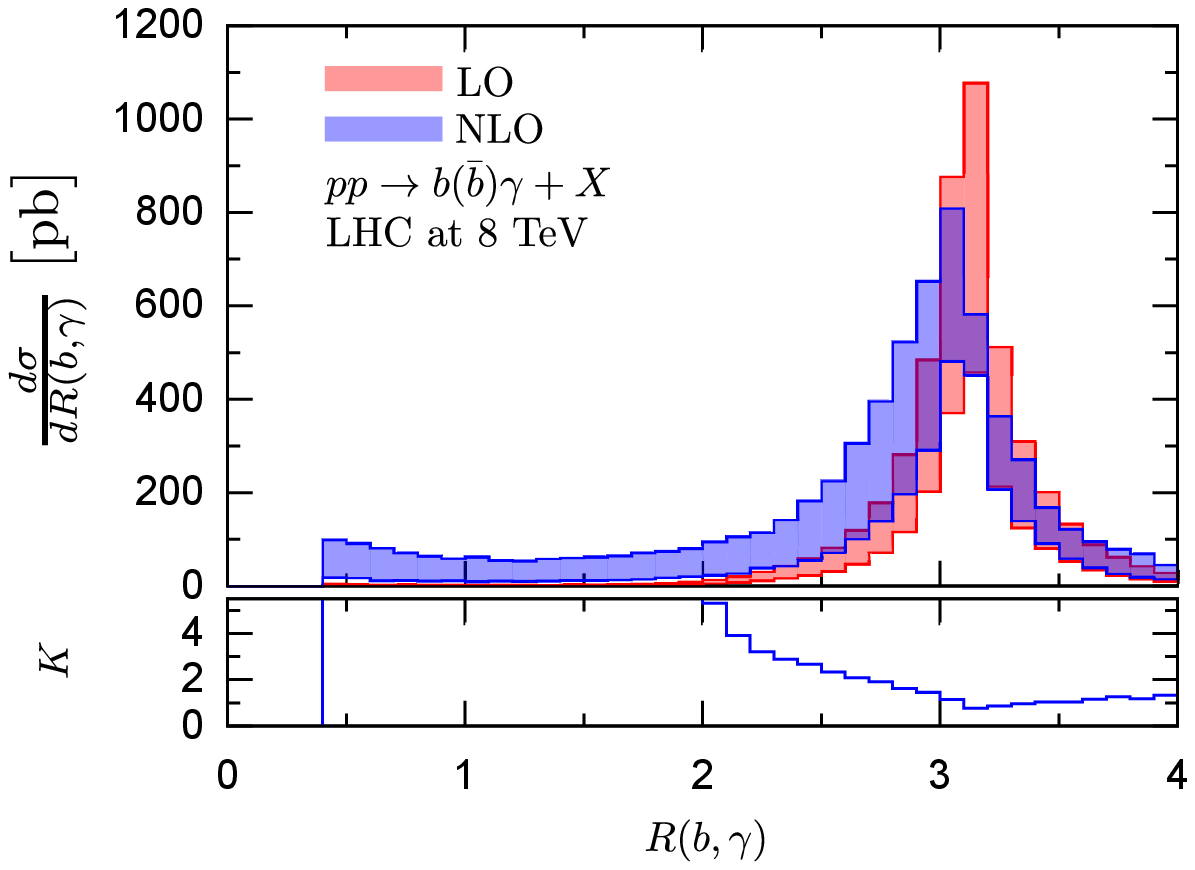}
\caption{The upper plots show the pseudorapidity distribution of the
  photon (left) and the separation between the $b$ jet and the photon
  (right) for $pp \rightarrow b(\bb)\gam+X$ (at least one $b$ jet
  identified in the final state) at the LHC with $\sqrt{s}=8$~TeV. The
  bands correspond to the variation of the renormalization and
  factorization scales in the interval $\mu_0/4 < \mu < 4 \mu_0$.  The
  lower plots show the bin-by-bin $K$-factor for the corresponding
  distributions.}
\label{fig:gam1b_dist2_lhc8}
\end{center}
\end{figure}

In Figs.~\ref{fig:gam1b_dist1_tev2}~and~\ref{fig:gam1b_dist2_tev2} we
show the photon and the $b$-jet transverse-momentum distribution, the
photon pseudorapidity and the photon-to-$b$-jet separation
distributions at the Tevatron. As for the 2$b$-jet case, the bands in
each figure represent the variation of the differential cross section
(bin-by-bin) when the renormalization and factorization scales
($\muR=\muF$) are varied in the range $\mu_0/4 \leq \mu \leq 4\mu_0$,
with $\mu_0=p_T(\gam)$, and the lower window of each figure gives the
bin-by-bin $K$-factor.  At the Tevatron, the $K$-factor for both the
$p_T(\gam)$ and $p_T(b)$ distributions decreases as we go from the
low- to the high-$p_T$ region, similar to what we observed in the
2$b$-tag case.  In
Figs.~\ref{fig:gam1b_dist1_lhc8}~and~\ref{fig:gam1b_dist2_lhc8}
analogous plots are presented for the LHC.  The impact of NLO QCD
corrections to the differential distributions are quite significant.
Similar to the 2$b$-tag case, the $K$-factor for the $p_T(\gam)$ and
$p_T(b)$ distributions grows as $p_T$ increases. The $R(\gam,b)$
distribution, however, does not exhibit the pinching that is present
in the 2$b$-tag case.

In Figs.~\ref{fig:gam1b_dist_tev2_sub} and
\ref{fig:gam1b_dist_lhc8_sub}, the contribution of the $q\qb$, $gg$,
and $qg$ channels to the NLO distributions are shown, allowing us to
argue that the strong scale dependence in the $p_T(\gam)$ and $p_T(b)$
distributions at the LHC are due to the $qg$ channel that contributes
significantly, almost as much as the $gg$ channel.  On the other hand,
at the Tevatron the scale dependence in the distributions shows quite
an improvement as $p_T(\gamma)$ or $p_T(b)$ grows, due to the $q\qb$
channel that dominates in the intermediate- to high-$p_T$ regions.

\begin{figure}
\begin{center}
\hspace{1.5cm}Fixed-cone isolation\hspace{4cm}Smooth-cone isolation \\
\includegraphics[width=7.5cm]{Gam1B_TEV02_pTgam_fixed_cone.eps}
\includegraphics[width=7.5cm]{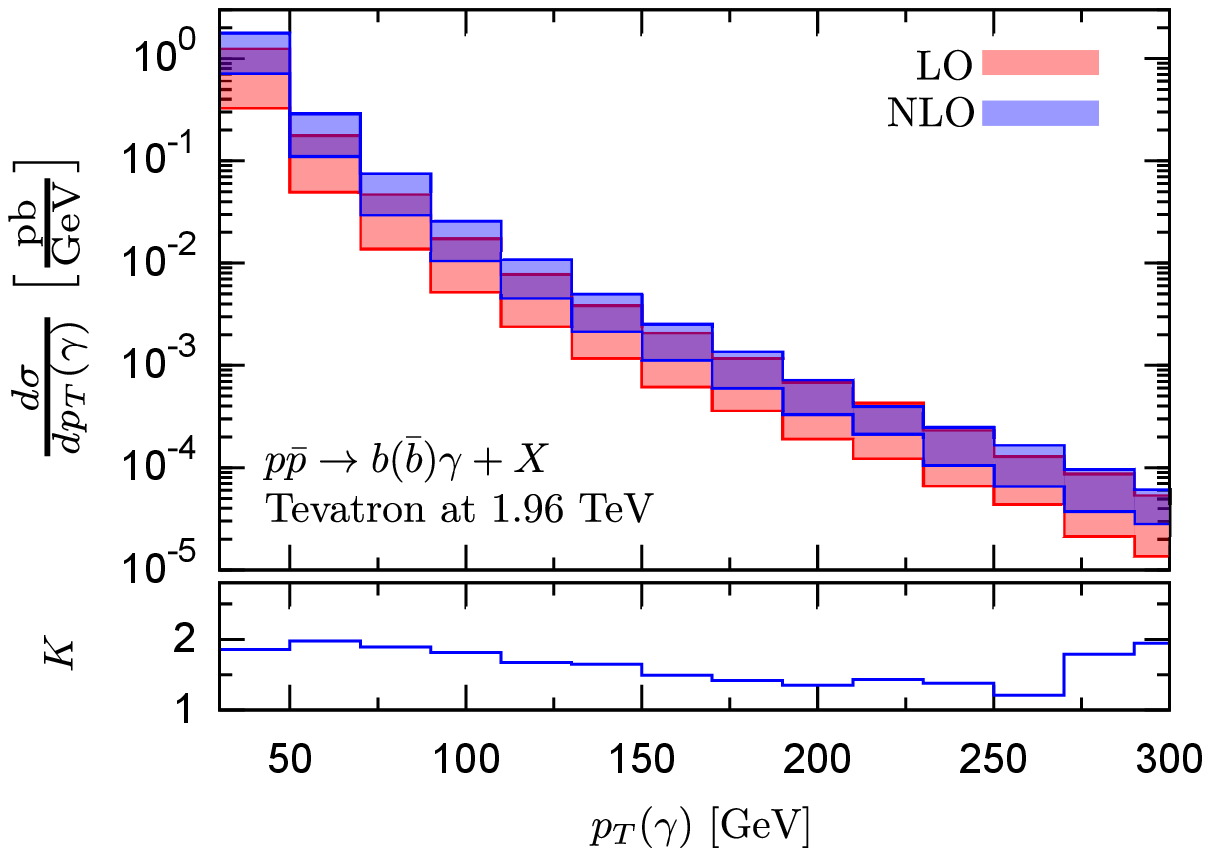}\\
\hspace{1.5cm}Fixed-cone isolation\hspace{4cm}Smooth-cone isolation \\
\includegraphics[width=7.5cm]{Gam1B_LHC08_pTgam_fixed_cone.eps}
\includegraphics[width=7.5cm]{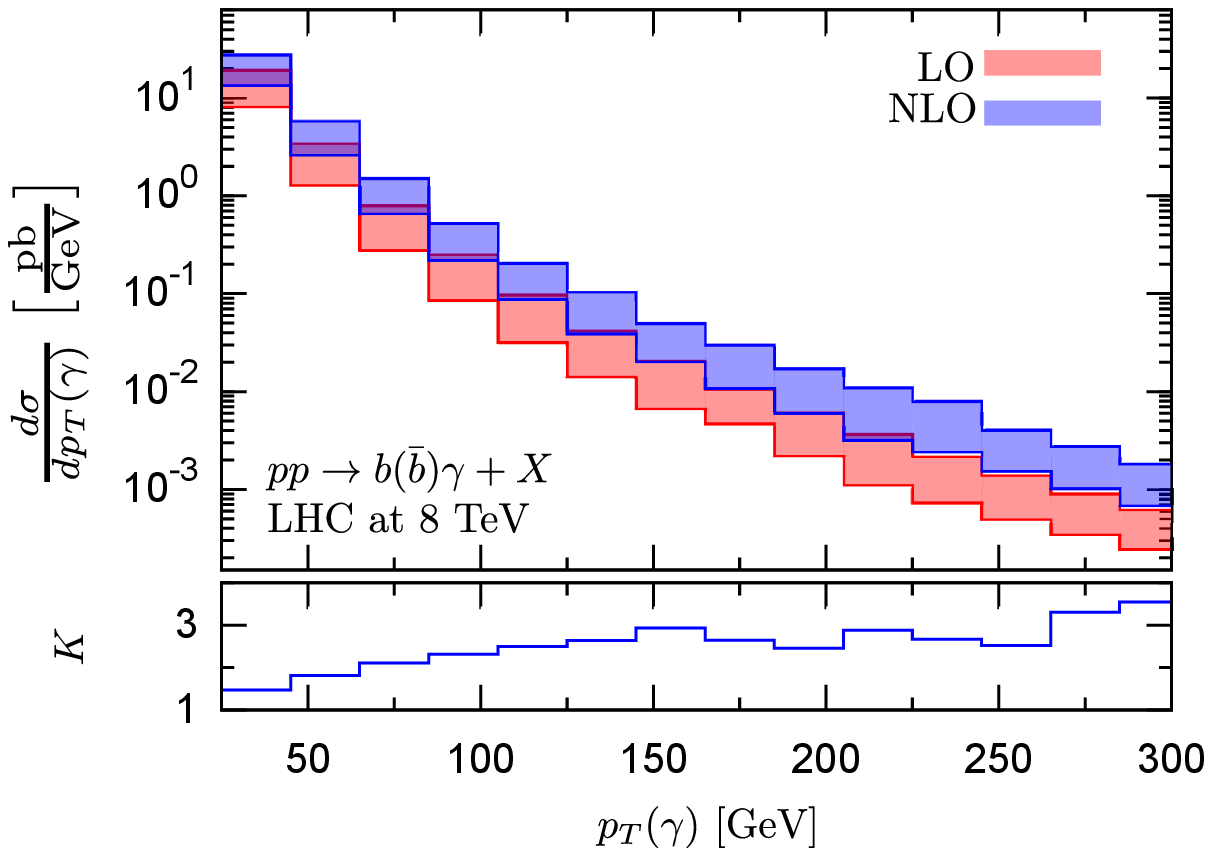}\\
\caption{Transverse-momentum distributions of the photon in $p\pb
  \rightarrow b(\bb)\gam+X$ (at least one $b$ jet identified in the
  final state) at the Tevatron with $\sqrt{s}=1.96$~TeV (top) and the
  LHC with $\sqrt{s}=8$~TeV (bottom), obtained using a fixed-cone
  (left) or a smooth-cone (right) isolation for the hard photon.  The
  right hand side plots are the same shown in
  Fig.~\ref{fig:gam1b_dist1_tev2} (Tevatron) and in
  Fig.~\ref{fig:gam1b_dist1_lhc8} (LHC) and they are repeated here for
  comparison at a glance. The bands correspond to the variation of
  the renormalization and factorization scales in the interval
  $\mu_0/4 < \mu < 4 \mu_0$.  The lower plots show the bin-by-bin
  $K$-factor for the corresponding distributions. }
\label{fig:gam1b_pt_gam_fixed_cone}
\end{center}
\end{figure}
Finally, in Figs.~\ref{fig:gam1b_pt_gam_fixed_cone} we show the
comparison between using a smooth-cone or fixed-cone isolation
algorithm for the hard photon. For the smooth-cone isolation prescription,
we use $R_0=0.4$ and $\epsilon=1$ (see Eq.~\ref{eq:frixiso}). 
We illustrate it both for the Tevatron and the LHC in
the case of the photon transverse-momentum distribution. The
difference is visible but small at the Tevatron and very small at the
LHC. 

\subsection{Comparison between 5FNS and 4FNS calculations}
\label{subsec:bbgam_4v5}

In Sec.~\ref{subsec:1b_4v5_theory}, we have discussed that the
predictions for direct photon production in association with a bottom
quark at hadron colliders, $pp(p\pb) \rightarrow \gamma+b+X$, can be
obtained from two different schemes. The first one is the FFS/4FNS
(4FNS in the following), where there is no initial-state bottom quark
in the partonic subprocesses and the mass of the bottom quark is
retained in the calculation.  This is essentially the $b\bb\gam$
calculation with at least one $b$ jet tagged in the final state, whose
results have been presented in the Sec.~\ref{subsec:bbgam_1b}. The
second scheme is the VFS/5FNS (5FNS in the following), where the
bottom quark is treated as massless and can be present in the initial
state of all contributing subprocesses. In this last scheme, the large
logarithms that appear due to the phase-space integration of the
unobserved final-state $b$ quark (only one $b$ quark is tagged) is
resummed and absorbed into the bottom-quark PDF. The calculation of
$pp(p\pb) \rightarrow \gamma+b+X$ in the 5FNS has been reported in
\cite{Stavreva:2009vi}, where a fixed-cone isolation is used for the
hard photon and the fragmentation component is included up to
$\oo(\alpha\alpha_s^2)$.

In this Section, we would like to compare the results for $pp(p\pb)
\rightarrow \gam+b+X$ from the two schemes at NLO QCD accuracy. In the
comparison the same hard-photon isolation prescription have to be
adopted. For this reason we have reproduced the 5FNS calculation
(details of which are given in Sec.~\ref{subsec:1b_4v5_theory}) 
using both a smooth-cone and
a fixed-cone isolation prescription. This last step has allowed us to
compare with Ref.~\cite{Stavreva:2009vi} and has also given us more
flexibility in the comparison with experimental data.

The one-loop amplitude for the $gb\rightarrow \gam b$ subprocess is
taken from Ref.~\cite{Aurenche:1986ff}, as implemented in the
MCFM~\cite{Campbell:2010ff} code.  The real corrections are computed
using the two-cutoff phase-space slicing method to extract the soft and collinear
singularities. The virtual+soft contribution to the $gb\rightarrow
\gam b$ subprocess also agrees with the result presented in
Ref.~\cite{Baer:1990ra}.  We have compared our 5FNS calculation with
the smooth-cone isolation prescription against a modified MCFM
code. Our $\gam +b+X$ implementation in MCFM is based on the
calculation of direct photon production at hadron colliders, $pp(p\pb)
\rightarrow \gamma+\mathrm{jet}+X$, modified by selecting the partonic
channels that are listed in Table~\ref{table:5FNSsubproc} with the
addition of imposing the suitable $b$-jet selection criteria.  We have
cross-checked our fixed-cone isolation results with the Authors of
Ref.~\cite{Stavreva:2009vi} and found consistency.  The only
difference with respect to Ref.~\cite{Stavreva:2009vi} is that we
include $\oo(\alpha\alpha_s)$ instead of $\oo(\alpha\alpha_s^2)$
fragmentation contributions, which can be argued to be indeed consistent
with the NLO QCD calculation and it is in any case expected to make a
minor difference since the whole point of the isolation procedure is
to reduce the impact of contributions from parton fragmentation (see
our discussion in Sec.~\ref{subsec:photon_isolation}).

\begin{figure}
\begin{center}
\includegraphics[width=7.5cm]{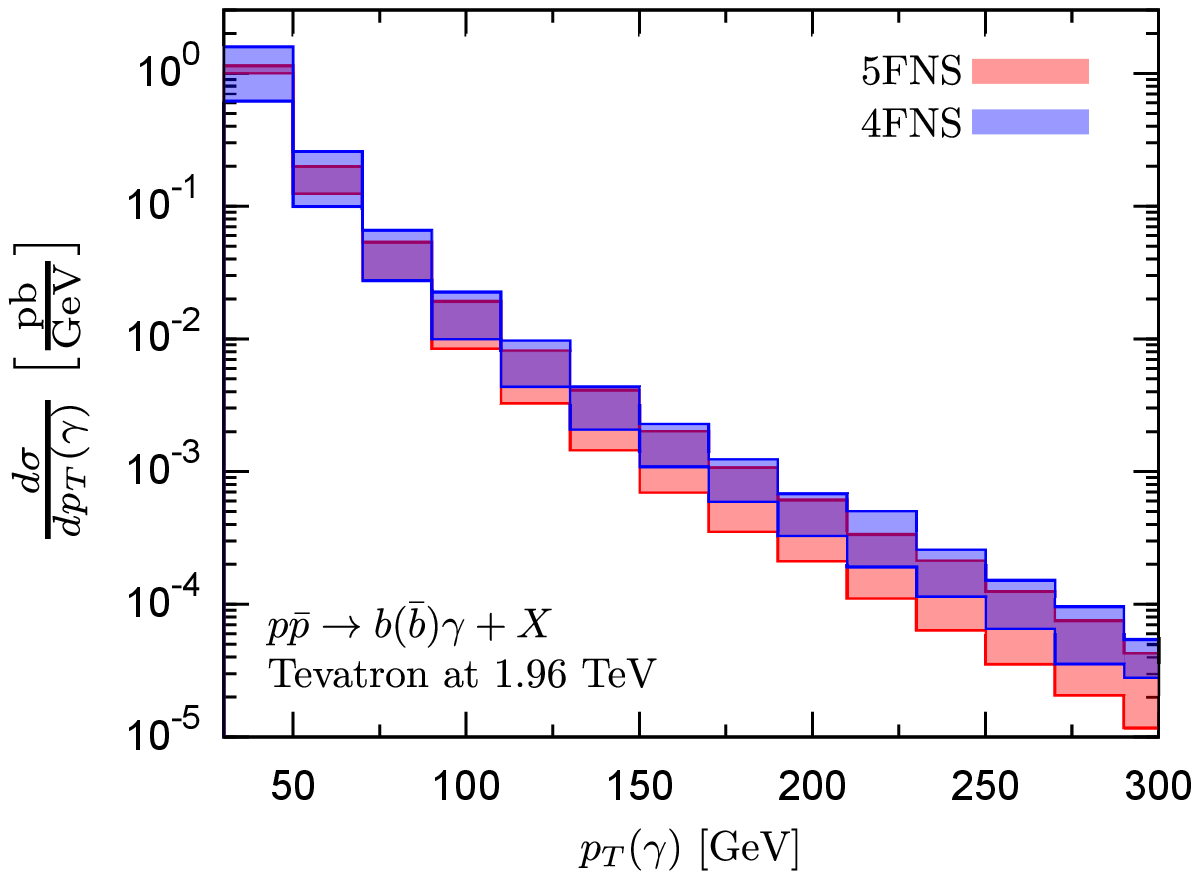}
\includegraphics[width=7.5cm]{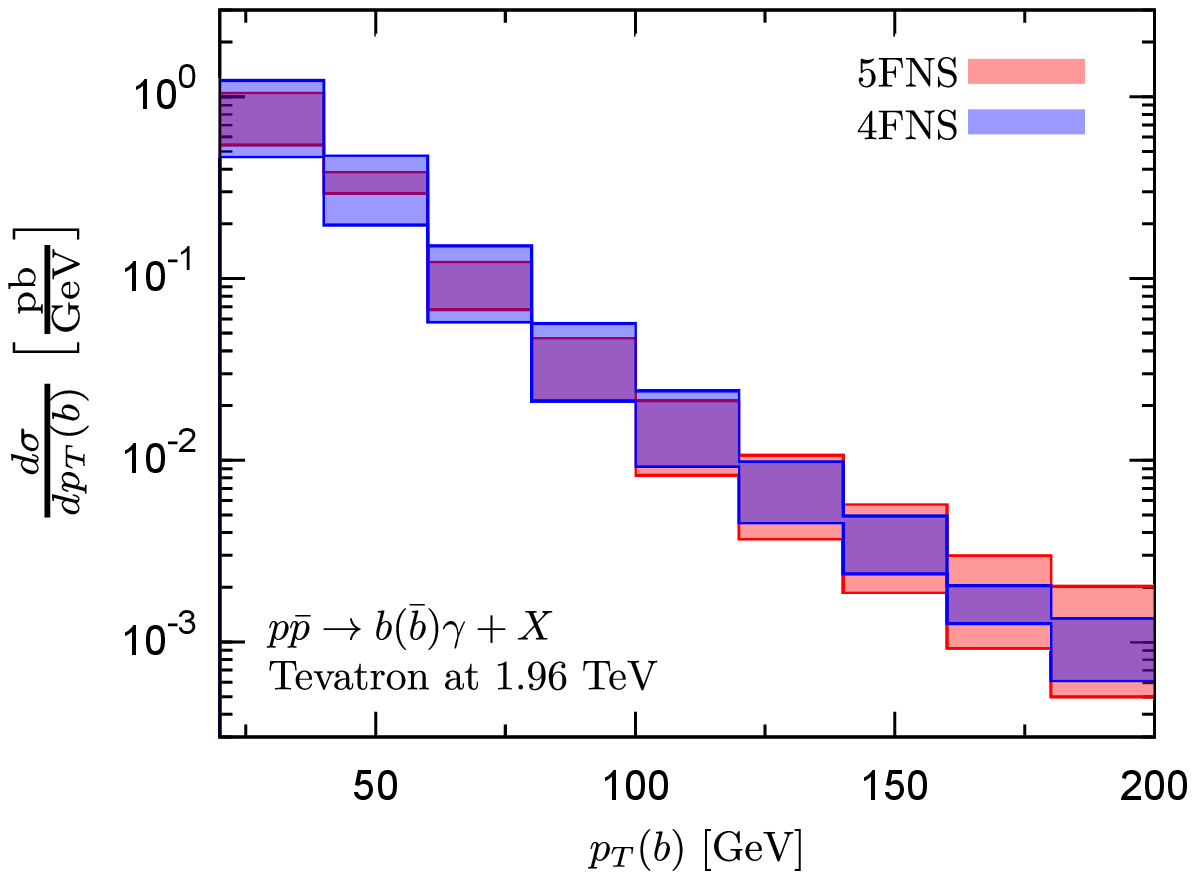}
\caption{Transverse-momentum distributions of the photon (left) and
  the $b$ jet (right) for $p\pb \rightarrow \gam+b+X$ (at least one
  $b$ jet identified in the final state) at the Tevatron with
  $\sqrt{s}=1.96$~TeV, obtained from the 4FNS (blue) and the 5FNS
  (red) calculations. The bands correspond to the variation of the
  renormalization and factorization scales in the interval $\mu_0/4 <
  \mu < 4 \mu_0$.}
\label{fig:gam1b_4v5_tev2}
\end{center}
\begin{center}
\includegraphics[width=7.5cm]{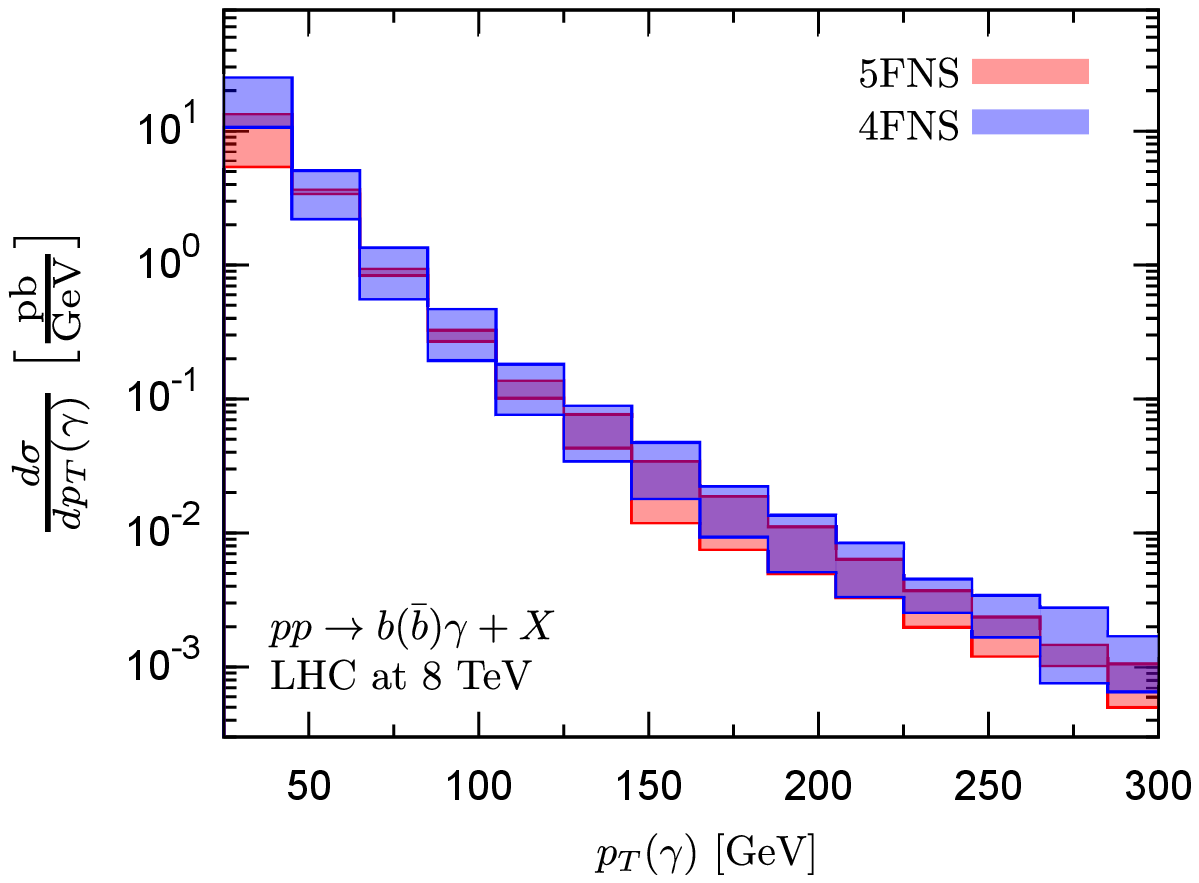} 
\includegraphics[width=7.5cm]{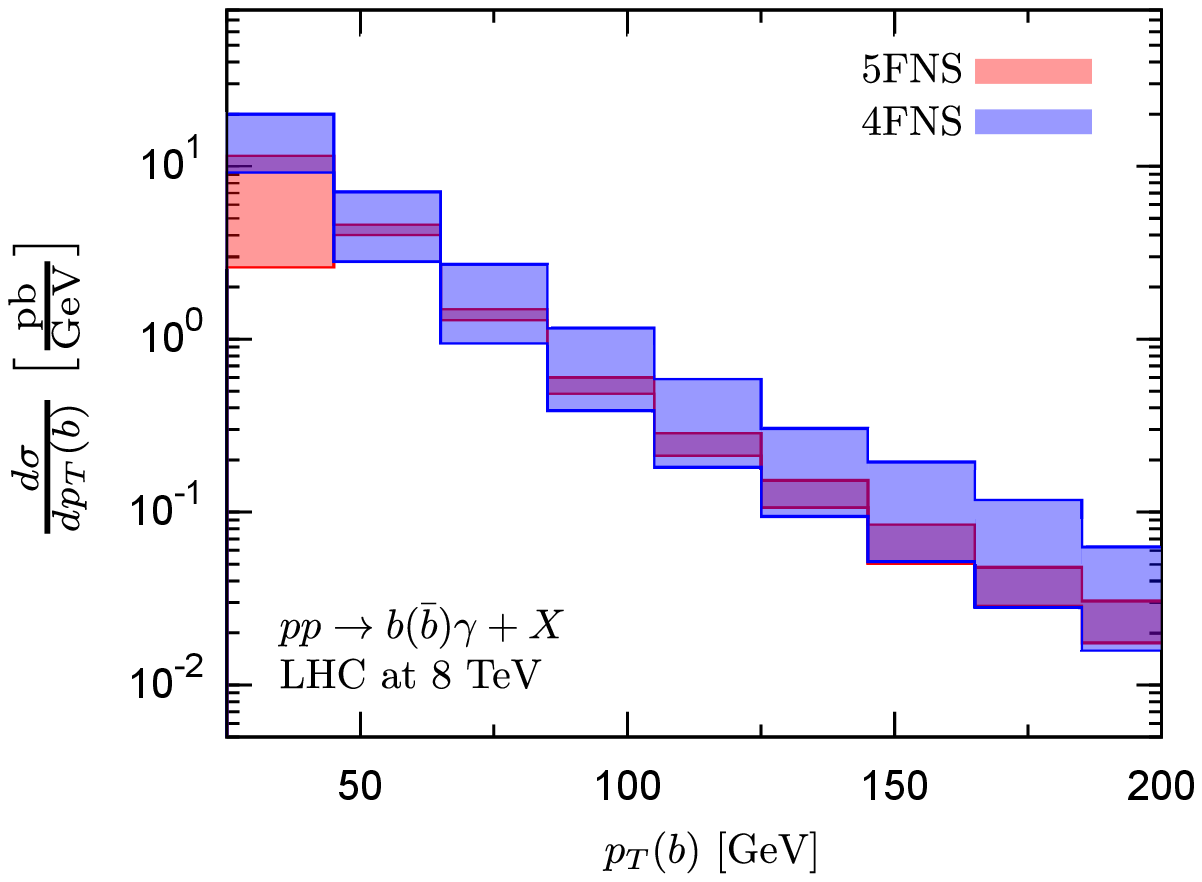}
\caption{Transverse-momentum distributions of the photon (left) and
  the $b$ jet (right) for $pp \rightarrow \gam+b+X$ (at least one $b$
  jet identified in the final state) at the LHC with $\sqrt{s}=8$~TeV,
  obtained from the 4FNS (blue) and the 5FNS (red) calculations. The
  bands correspond to the variation of the renormalization and
  factorization scales in the interval $\mu_0/4 < \mu < 4 \mu_0$.}
\label{fig:gam1b_4v5_lhc}
\end{center}
\end{figure}

In Figs.~\ref{fig:gam1b_4v5_lhc}~and~\ref{fig:gam1b_4v5_tev2} we show
the comparison between the 4FNS and 5FNS calculations for the photon
and the $b$-jet transverse-momentum distributions at the Tevatron and
at the LHC, using a fixed-cone isolation prescription. The comparison
using a smooth-cone isolation prescription leads to very similar
results.  From the plots one can see that both at the Tevatron and at
the LHC the 4FNS predictions lead the 5FNS ones, but they are overall
compatible within their theoretical systematic uncertainty which we
base only on the renormalization and factorization scale
uncertainty. This is indeed what is expected from a relatively
well-behaved QCD perturbative series. It also indicates that the
process under consideration is moderately sensitive to the kind of
kinematic logarithms that are resummed in the $b$-quark PDF.

It is however interesting to understand the behavior of the
scale-variation band, which is indeed different for the Tevatron and
the LHC.  At the LHC the distributions from the 5FNS calculation have
smaller scale-uncertainty bands compared to the 4FNS counterparts.  As
it has been discussed in \cite{Stavreva:2009vi}, at the LHC, the
dominant contribution to the $p_T(\gam)$ distribution comes from the
$Qg \rightarrow \gam Q$, $gg \rightarrow \gam b\bb$, and $Qg
\rightarrow \gam Q g$ channels (see Table~\ref{table:5FNSsubproc}),
where potentially large kinematic logarithms have been resummed in the
$b$-quark PDF, resulting in the better scale-dependence behavior.  On
the other hand, at the Tevatron, while the scale-variation bands from
the 5FNS calculation are smaller than the 4FNS calculation at
low $p_T(\gam)$, as $p_T(\gam)$ increases the 5FNS bands are getting larger,
while the 4FNS bands get sensibly smaller. As we have seen in
Fig.~\ref{fig:gam1b_dist_tev2_sub}, the $q\qb \rightarrow b\bb\gam$
channel dominates in the intermediate- to high-$p_T$ region, and since
this piece of the calculation enters in the 5FNS real correction as a
tree level process, it still has a strong scale dependence.

\subsection{Comparison of NLO QCD predictions for $\gam+b+X$ 
 with Tevatron data}
\label{subsec:1bgam_vs_tevdata}

The calculation of $b\bb\gam$ production at NLO QCD accuracy allows us
to improve the theoretical prediction of the $pp(p\pb) \rightarrow
\gam +b +X$ process at hadron colliders and to compare with the recent
measurements of direct-photon production with at least one $b$ jet
($\gamma+b+X$) at the Tevatron by the CDF \cite{Aaltonen:2013coa} and
D0~\cite{Abazov:2012ea} collaborations.

Since the measurement of the $\gam+b+X$ process at the Tevatron
employed the fixed-cone photon isolation we will specify both 4FNS and
5FNS to this prescription and adopt otherwise the identification cuts
used by CDF~\cite{Aaltonen:2013coa},
\begin{center}
\begin{tabular} { l l}
$p_T(\gam) > 30$~GeV, $|\eta(\gam)| < 1$\,\,\,,\\ 
Photon isolation: $R_0=0.4$, $E_T^{\mathrm{max}}=2$~GeV\,\,\,,\\
$p_T(b,j) > 20$~GeV, $|\eta(b,j)| < 1.5$, $R=0.4$\,\,\,,
\end{tabular}
\end{center}
and D0~\cite{Abazov:2012ea},
\begin{center}
\begin{tabular} { l l}
$p_T(\gam) > 30$~GeV, $|\eta(\gam)| < 1$\,\,\,\,\mbox{and}\,\,\,\,
$1.5<|\eta(\gam)| < 2.5$\,\,\,,\\ 
Photon isolation: $R_0=0.4$, $E_T^{\mathrm{max}}=2.5$~GeV\,\,\,,\\
$p_T(b,j) > 15$~GeV, $|\eta(b,j)| < 1.5$, $R=0.5$\,\,\,,
\end{tabular}
\end{center}
when comparing with each experiment respectively. From the detailed
tables given by both CDF~\cite{Aaltonen:2013coa} and
D0~\cite{Abazov:2012ea} in their papers we have been able to provide a
comparison between the experimental data, the 5FNS NLO predictions of
Ref.~\cite{Stavreva:2009vi}, and our 4FNS NLO predictions, including
the corresponding statistical and systematic uncertainties.

Since the results of Ref.~\cite{Stavreva:2009vi} used CTEQ6.6 as PDF
set, and estimate the systematic theoretical uncertainty from
renormalization- and factorization-scale dependence by varying them in
the $\mu_0/2 < \muR=\muF < 2 \mu_0$ range ($\mu_0=p_T(\gam)$), we
ought to specify our results to this set-up to have a more adequate
comparison\footnote{For the case of the comparison with D0 in the
  forward region we only plot 5FNS for $\mu=\mu_0$ because we were not
  provided the full scale variation. We decided not to reproduce the
  band ourselves because the results of Ref.~\cite{Stavreva:2009vi},
  to which the experiments have compared in their papers, include
  fragmentation contributions at $O(\alpha\alpha_s^2)$ while we only
  include them at $O(\alpha\alpha_s)$. These effects are small, but we
  prefer to show a consistent set of 5FNS results.}.  At the same
time, we think that it is important to have a comparison using a more
up-to-date set of PDFs, such as the CT10nlo\_nf4 set that we use in
the rest of this paper. We also consider the variation of $\mu_R$ and
$\mu_F$ in the $\mu_0/4 < \muR=\muF < 4 \mu_0$ range a more accurate
representation of the theoretical uncertainty of our predictions,
given what has been discussed in Sec.~\ref{subsec:bbgam_1b}.
Therefore, for each experiment we will present a comparison using the
same setup (CTEQ6.6 and $\mu_0/2 < \muR=\muF < 2 \mu_0$) for both 5FNS
and 4FNS NLO results, and a comparison where the 5FNS results are the
same but the 4FNS results use updated PDFs and a more realistic
systematic uncertainty (CT10nlo\_nf4 and $\mu_0/4 < \muR=\muF < 4
\mu_0$).

\begin{figure}
\begin{center}
\includegraphics[width=7cm]{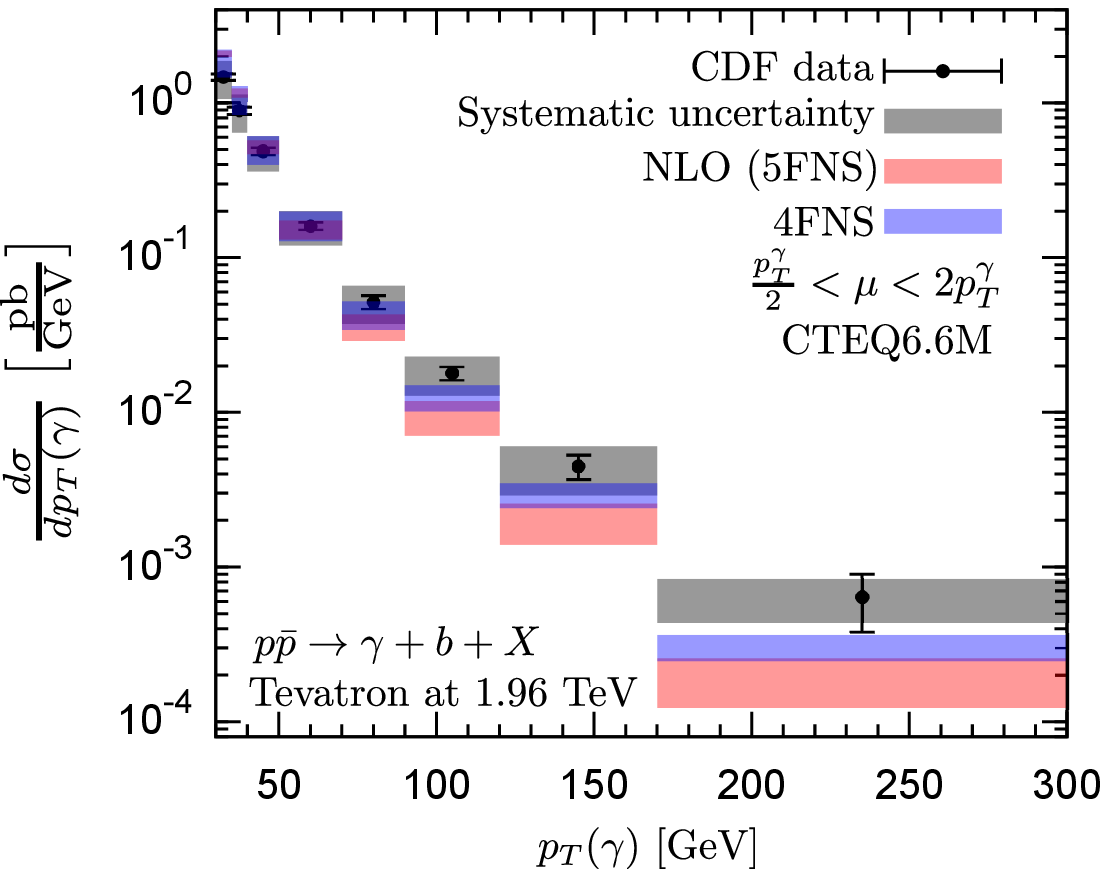} 
\includegraphics[width=7cm]{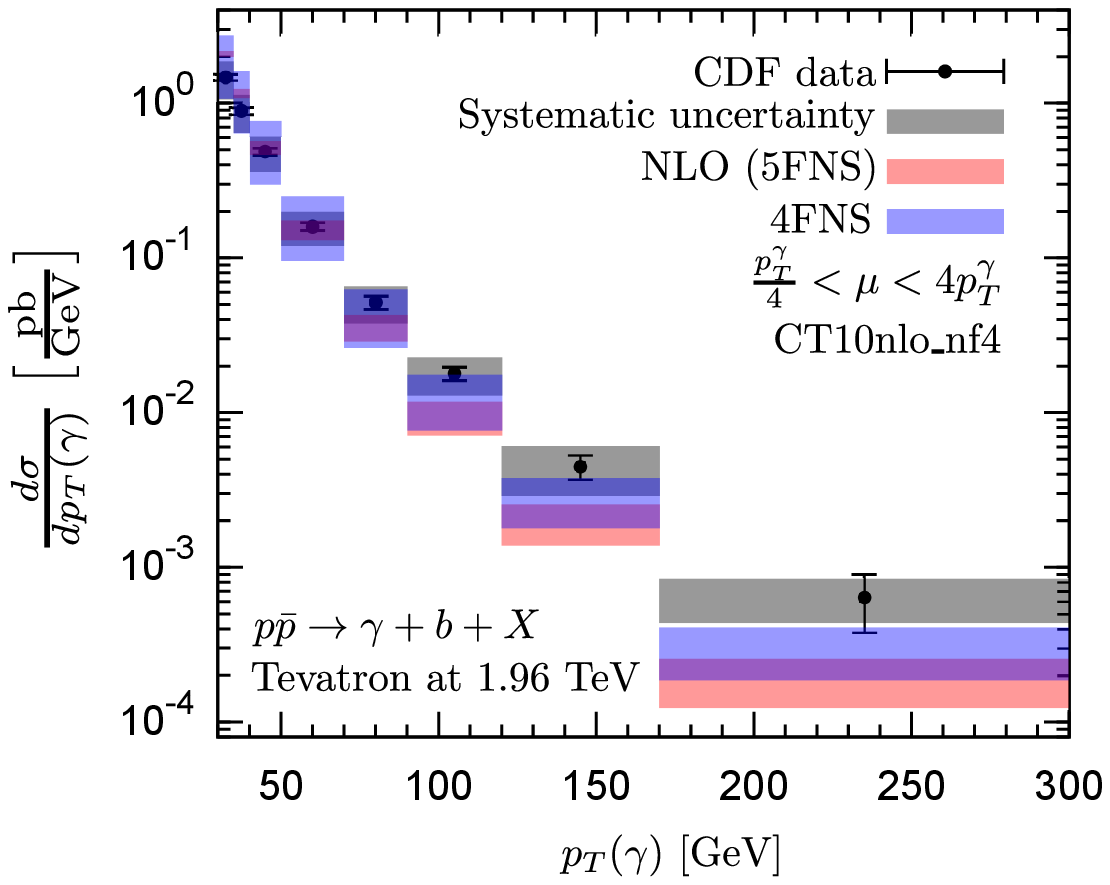} 
\caption{The photon-$p_T$ distribution for $\gamma+b+X$ at the
  Tevatron with $\sqrt{s}=1.96$~TeV and the identification cuts used
  by the CDF experiment.  The numbers for the CDF data and
  the NLO 5FNS calculation (red)~\cite{Stavreva:2009vi} are taken from
  \cite{Aaltonen:2013coa}.  The NLO 4FNS results (blue) are from this
  paper.  For the experimental data, the gray band represent the
  systematic uncertainty only, while both systematic and statistic
  uncertainties are included in the error bars.  The 5FNS NLO results
  have been obtained using CTEQ6.6 PDFs and the systematic
  uncertainty band corresponds to varying $\muR$ and $\muF$ in the $\mu_0/2 <
  \muR=\muF < 2 \mu_0$ range.  The 4FNS results have been obtained
  using the same set-up (left) as well as using CT10nlo\_nf4 PDFs and
  varying $\muR$ and $\muF$ in the $\mu_0/4 < \muR=\muF < 4 \mu_0$
  range (right).}
\label{fig:CDFvNLO}
\end{center}
\end{figure}

\begin{figure}
\begin{center}
\includegraphics[width=7cm]{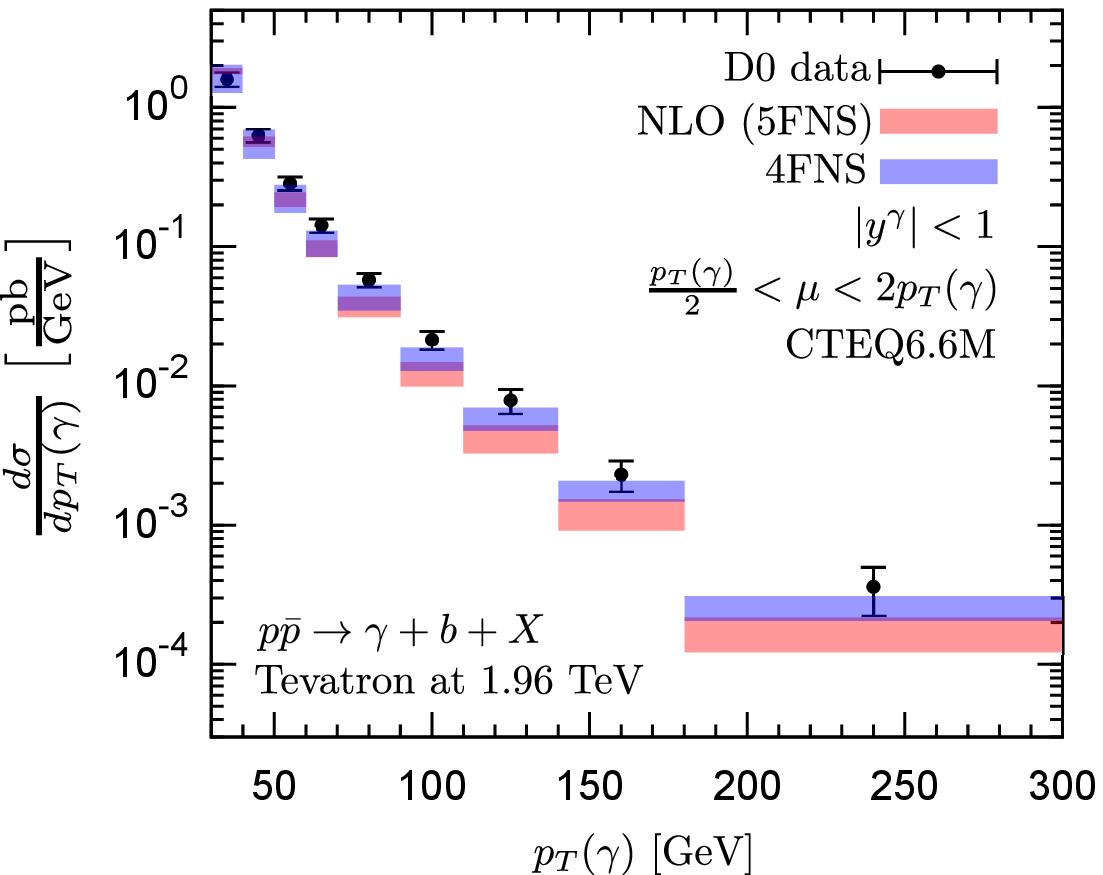} 
\includegraphics[width=7cm]{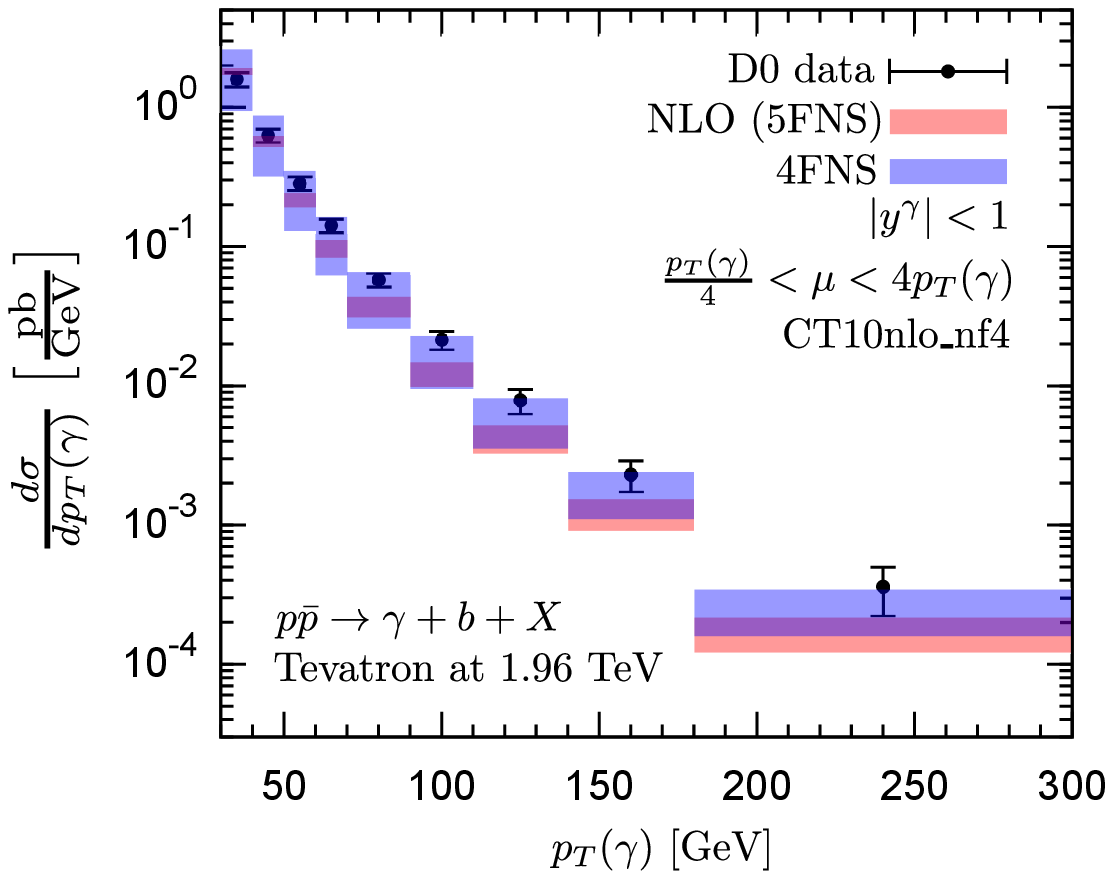}
\\
\includegraphics[width=7cm]{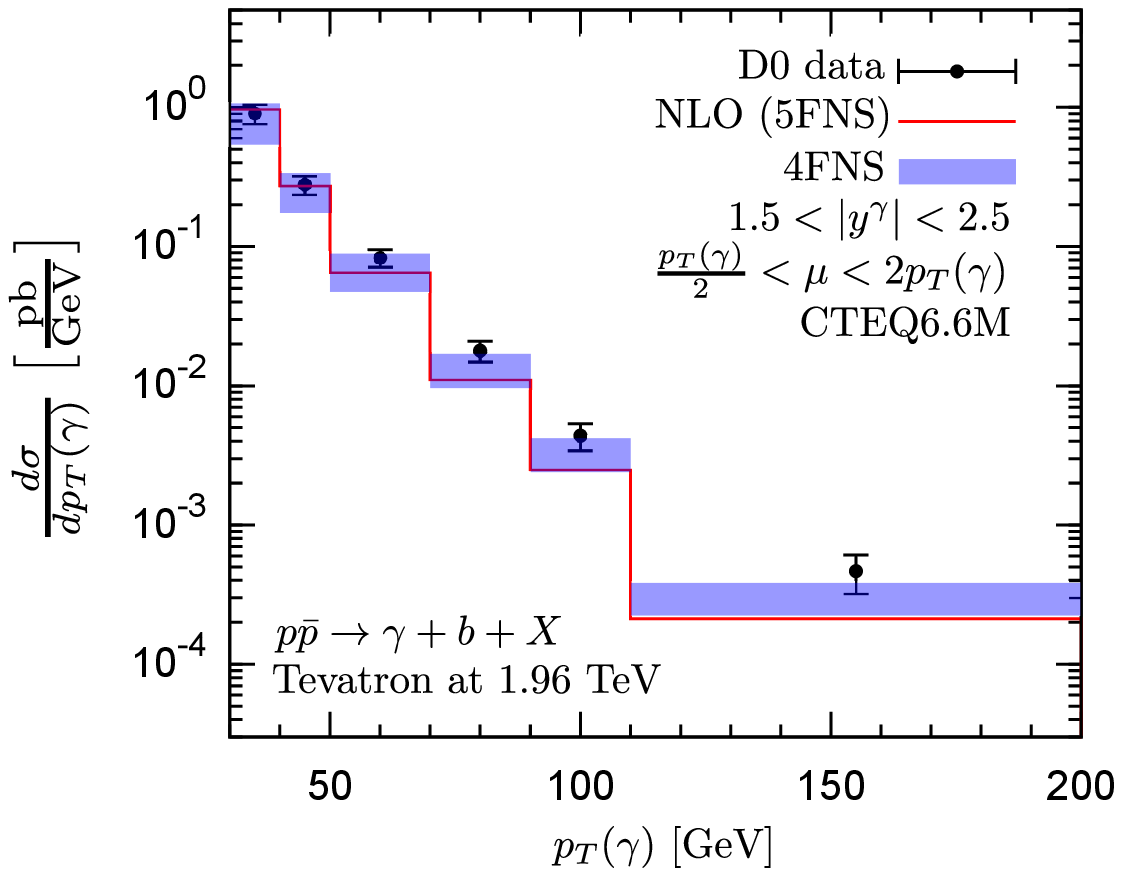} 
\includegraphics[width=7cm]{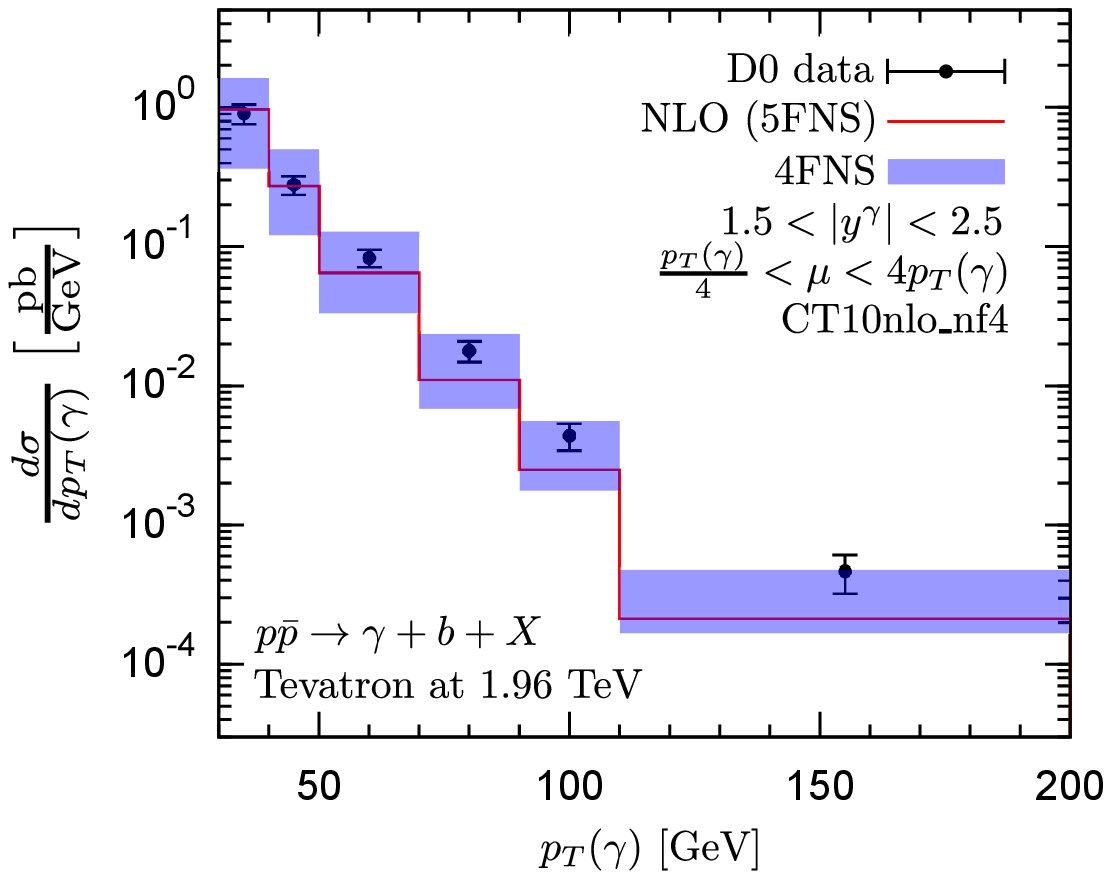}
\caption{The photon-$p_T$ distribution for $\gamma+b+X$ at the
  Tevatron with $\sqrt{s}=1.96$~TeV and the identification cuts used
  by the D0 experiment.  The numbers for the CDF data and the NLO 5FNS
  calculation (red)~\cite{Stavreva:2009vi} are taken from
  \cite{Abazov:2012ea}.  The NLO 4FNS results (blue) are from this
  paper.  For the experimental data, we have considered both the
  central- and the forward-rapidity data. Both systematic and
  statistic uncertainties are included in the error bars.  The 5FNS
  NLO results have been obtained using CTEQ6.6 PDFs and
  uncertainty band corresponds to varying $\muR$ and $\muF$ in the
  $\mu_0/2 < \muR=\muF < 2 \mu_0$ range.  For the forward-rapidity
  region only the central values were available.  The 4FNS results
  have been obtained using the same set-up (left) as well as using
  CT10nlo\_nf4 PDFs and varying $\muR$ and $\muF$ in the $\mu_0/4 <
  \muR=\muF < 4 \mu_0$ range (right).}
\label{fig:D0vNLO}
\end{center}
\end{figure}

The results are presented in Figs.~\ref{fig:CDFvNLO} and
\ref{fig:D0vNLO} for CDF and D0 respectively. We notice that the 4FNS
NLO results show a better agreement with data in the high $p_T(\gam)$
region. This is not surprising since it is mainly induced by the fact
that at the Tevatron $q\qb\rightarrow b\bb\gam$ is the leading
subprocess in the high-$p_T(\gam)$ region and this channel is only
included at tree level in the 5FNS calculation, while it is enhanced
by NLO corrections in the 4FNS calculation.


\section{Conclusions}
\label{sec:conclusions}

In this paper we have reported the NLO QCD calculation of
hard-photon production with heavy quarks at hadron colliders
($pp(p\pb)\rightarrow Q\bar{Q}\gam$ with $Q=t,b$), and we have
provided results for $\gam +2b$- and $\gam +1b$-jet production at both
the Tevatron with center-of-mass energy 1.96~TeV and the Large Hadron
Collider with center-of-mass energy 8~TeV. We have compared with the
CDF and D0 measurements of hard-photon production with at least one
$b$ jet~\cite{Abazov:2012ea,Aaltonen:2013coa} and found consistency.
Future results for $\gam +1b$- and $\gam +2b$-jet production from
ATLAS and CMS will offer the opportunity to test the NLO QCD
prediction for these processes in different energy and kinematic
regimes. Future studies could also make use of the existing
implementation of our calculation in a full-fledged NLO parton-shower
Monte Carlo (e.g. Sherpa).

The level of uncertainty in the theoretical predictions is adequate to
compare with current experimental accuracies, and offer the first
opportunity to constrain the $b$-quark parton distribution function.

Finally, a non trivial extension of our calculation could involve
considering the production of a hard photon with charm quarks to be
used as a direct probe of the charm-quark parton distribution
function.

\section*{Acknowledgements}
H.~B.~H. and L.~R. would like to greatly thank Tzvetalina Stavreva
for her help in comparing with the results of
Ref.~\cite{Stavreva:2009vi}, Stefan Hoeche for his help in
implementing the NLO calculation of $Q\bar{Q}\gamma$ into the
Sherpa Monte Carlo event generator, Ciaran Williams and Nobuo Sato 
for useful discussions.
L.~R. would like to thank the Aspen Center for Physics for its kind
hospitality while part of this work was being completed.
The work of H.~B.~H. and L.~R.~is supported in part by
the U.S. Department of Energy under grants DE-FG02-13ER41942.
\bibliography{bbgamma}

\end{document}